\newcommand{\CLASSINPUTtoptextmargin}{19mm}
\newcommand{\CLASSINPUTbottomtextmargin}{18mm} 
\documentclass[10pt,conference,final, a4paper]{ieeetran}

\usepackage[utf8]{inputenc}

\usepackage[sorting=none]{biblatex}
\usepackage{qcircuit}
\usepackage{braket}
\usepackage{amsmath,amssymb,amsmath} 
\usepackage{multirow} 
\usepackage{booktabs} 
\usepackage{graphicx} 
\usepackage{enumitem} 
\usepackage{caption} 
\usepackage{subcaption} 
\usepackage[linesnumbered,ruled,vlined]{algorithm2e}
\usepackage{listings} 
\usepackage{caption} 
\usepackage{subcaption} 
\usepackage{color}
\usepackage{hyperref}
\usepackage{textcomp}
\usepackage{algorithmic}
\usepackage{siunitx} 
\usepackage{bm}

\usepackage{wasysym} 
\newcommand{\controlxo}{ *+<.01em>{\LEFTcircle}}
\newcommand{\ctrlxo}[1]{\controlxo \qwx[#1] \qw} 
\newcommand{\qwdash}[1][-1]{\ar @{.} [0,#1]}

\DeclareMathSymbol{\shortminus}{\mathbin}{AMSa}{"39}

\usepackage{chngcntr} 
\usepackage{newfloat}

\DeclareFloatingEnvironment[
  fileext = los ,
  listname = {List of circuits} ,
  name = Circuit
]{circuit}

\usepackage[capitalise]{cleveref}
\crefname{circuit}{Circ.}{Circs.} 
\Crefname{circuit}{Circuit}{Circuits} 

\newif\ifdraft
\drafttrue
\ifdraft
\definecolor{teal}{rgb}{0,0.5,0.5}
\definecolor{violet}{rgb}{0.5,0,0.5}
\newcommand{\note}[1]{ {\textcolor{blue} { **: #1 }}}
\newcommand{\alnote}[1]{ {\textcolor{red} { ***Andre: #1 }}}
\newcommand{\annenote}[1]{ {\textcolor{violet} { ***Anneriet: #1 }}}
\newcommand{\marvnote}[1]{ {\textcolor{blue} { ***Marvin: #1 }}}
\newcommand{\ewannote}[1]{ {\textcolor{cyan} { ***Ewan: #1 }}}
\newcommand{\zaidnote}[1]{ {\textcolor{teal} { ***Zaid: #1 }}}
\else
\newcommand{\note}[1]{}
\newcommand{\alnote}[1]{}
\newcommand{\annenote}[1]{}
\newcommand{\marvnote}[1]{}
\newcommand{\ewannote}[1]{}
\newcommand{\zaidnote}[1]{}
\fi

\author{\IEEEauthorblockN{Anna M. Krol\href{https://orcid.org/0000-0003-0066-4299}{\includegraphics[scale=0.0035]{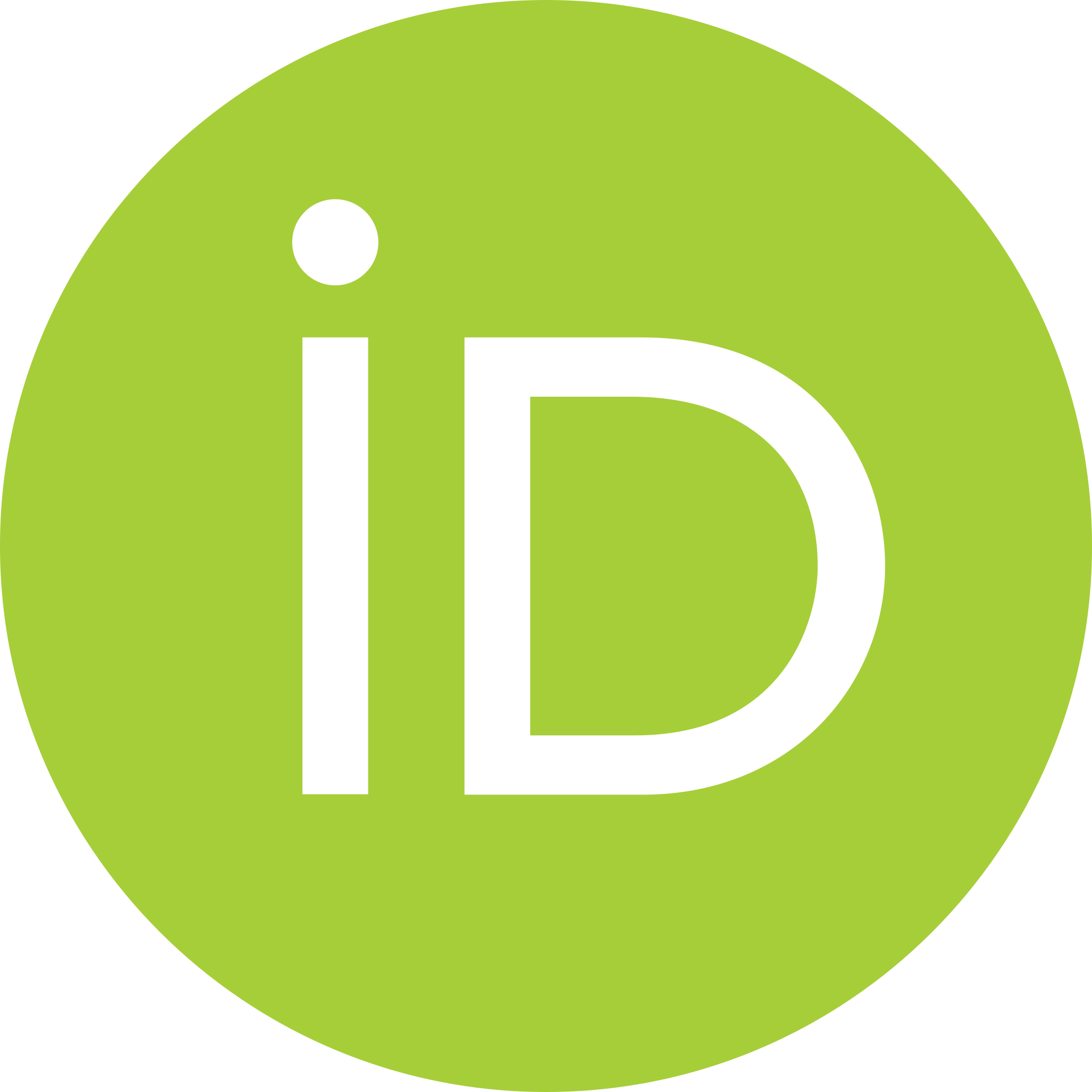}}\IEEEauthorrefmark{1}, Marvin Erdmann\IEEEauthorrefmark{2}, Ewan Munro\IEEEauthorrefmark{3}, Andre Luckow\IEEEauthorrefmark{2,}\IEEEauthorrefmark{4}, Zaid Al-Ars\IEEEauthorrefmark{1}}
\IEEEauthorblockA{\IEEEauthorrefmark{1}Delft University of Technology, Delft, The Netherlands}
\IEEEauthorblockA{\IEEEauthorrefmark{2}BMW Group, Munich, Germany}
\IEEEauthorblockA{\IEEEauthorrefmark{3}Entropica Labs, Singapore}
\IEEEauthorblockA{\IEEEauthorrefmark{4}Ludwig Maximilian University, Munich, Germany}
}

\begin{document}
\begin{refsection}
\title{Assessing the Requirements for Industry Relevant Quantum Computation}

\maketitle
\begin{abstract}
In this paper, we use open-source tools to perform quantum resource estimation to assess the requirements for industry-relevant quantum computation. Our analysis uses the problem of industrial shift scheduling in manufacturing and the Quantum Industrial Shift Scheduling algorithm. We base our figures of merit on current technology, as well as theoretical high-fidelity scenarios for superconducting qubit platforms. We find that the execution time of gate and measurement operations determines the overall computational runtime more strongly than the system error rates. Moreover, achieving a quantum speedup would not only require low system error rates ($\bm{10^{-6}}$ or better), but also measurement operations with an execution time below $\bm{10}\mskip\thinmuskip$ns. This rules out the possibility of near-term quantum advantage for this use case, and suggests that significant technological or algorithmic progress will be needed before such an advantage can be achieved.

\end{abstract}

\section{Introduction}
Industrial shift scheduling is an essential part of efficiently planning and running operations in the manufacturing sector. The challenge is to find the optimal production schedule for an end-to-end manufacturing system with multiple production sites. This schedule must comply with numerous constraints, including legal regulations and limited intermediate storage between the production sites. In volume-intensive industry sectors such as the automotive industry, one must additionally meet a production target corridor. The optimization goal is to minimize labor costs while satisfying all constraints.

The Quantum algorithm for Industrial Shift Scheduling (QISS)~\cite{art:krol2024qiss} provides the first fully quantum approach to finding exact solutions to volume-constrained industrial labor planning problems. Based on Grover Adaptive Search (GAS)~\cite{art:PASwithgrovers,art:GASforCPBO}, it inherits the asymptotic quadratic speedup of Grover's algorithm over classical unstructured search methods such as brute-force or random search. 

However, the problem size at which this quadratic speedup leads to a practical speedup is subject to limitations. On the one hand, it is impractical to seek exact solutions for very large problems, because: 1) the solution space grows exponentially with the problem size; and 2) the constraints typically impose very little structure on the solution space. Hence, one must resort to (classical) heuristic approaches, such as simulated annealing~\cite{art:Kirkpatrick1983} or tensor network methods~\cite{banner2023quantum}. On the other hand, for sufficiently small problems where finding exact solutions may be achievable, the inferior clock speed of quantum computers compared to classical computers will tend to wash out the quadratic speedup~\cite{art:TroyerBeyondQuadratic}. A natural question is then: does a regime exist where QISS can return exact solutions with a runtime that is both acceptable in a real-world setting, and superior to that of classical unstructured search? 

In this paper, we investigate this question systematically by estimating the resources required for the execution of QISS (described in \cref{QISS_section} below) on a fault-tolerant quantum computer using the surface code \cite{art:surfacecodes} for quantum error correction, as a function of problem size. Through comparison with the runtime of classical unstructured search, we evaluate the prospects for a practical speedup using QISS under various scenarios, each with different assumptions about the characteristics of the quantum computer. Using open-source resource estimation tools, we find the resources required for a speedup to be highly demanding, far out of reach for all existing and planned technology. By exploring different parameter scenarios, we quantify the role of key metrics such as qubit error rates and operation execution times in the quest for achieving a speedup.

In terms of the number of qubits, we find that currently planned quantum computers (i.e.\ those illustrated in \cref{fig:quantumroadmap}) will not be large enough to tackle instances of the shift scheduling problem where a classical unstructured search of the solution space becomes infeasible. This is broadly in line with the conclusions of other recent work to assess the required resources for quantum advantage~\cite{art:TroyerBeyondQuadratic,art:resourceestpaper,art:factor2028bitrsa,art:usingazureresest}.

The paper is organized as follows. Background information about current quantum computing platforms, quantum error correction and the QISS algorithm is given in \cref{sec:background}. An overview of the resource estimation process 
is given in \cref{sec:resourceest}. \cref{sec:currentscens} assesses the resource requirements for the execution of QISS using quantum computers that may be accessible in the near-to-mid~term, while \cref{sec:futurescens} investigates the prospects for a speedup using idealized technology with fast, high-fidelity operations. Our conclusions and outlook can be found in \cref{sec:conclusion}.

\section{Background} 
\label{sec:background}
In this section, we give background information on current and planned quantum computing platforms, quantum error correction codes, and we provide a brief overview of our use case and the QISS algorithm.

\subsection{Quantum computing technologies}

Quantum computers are being developed by numerous academic, governmental, and industrial organizations worldwide today, covering several different types of qubit technologies. The key figures of merit characterizing these systems include: the total number of qubits, the limiting physical error rate, the degree of qubit connectivity, and the time required to execute physical operations such as quantum gates and measurements. Notably, there is currently no single platform type or device that leads across all such figures of merit.

\begin{figure}[hbt]
    \centering
    \includegraphics[width=\linewidth]{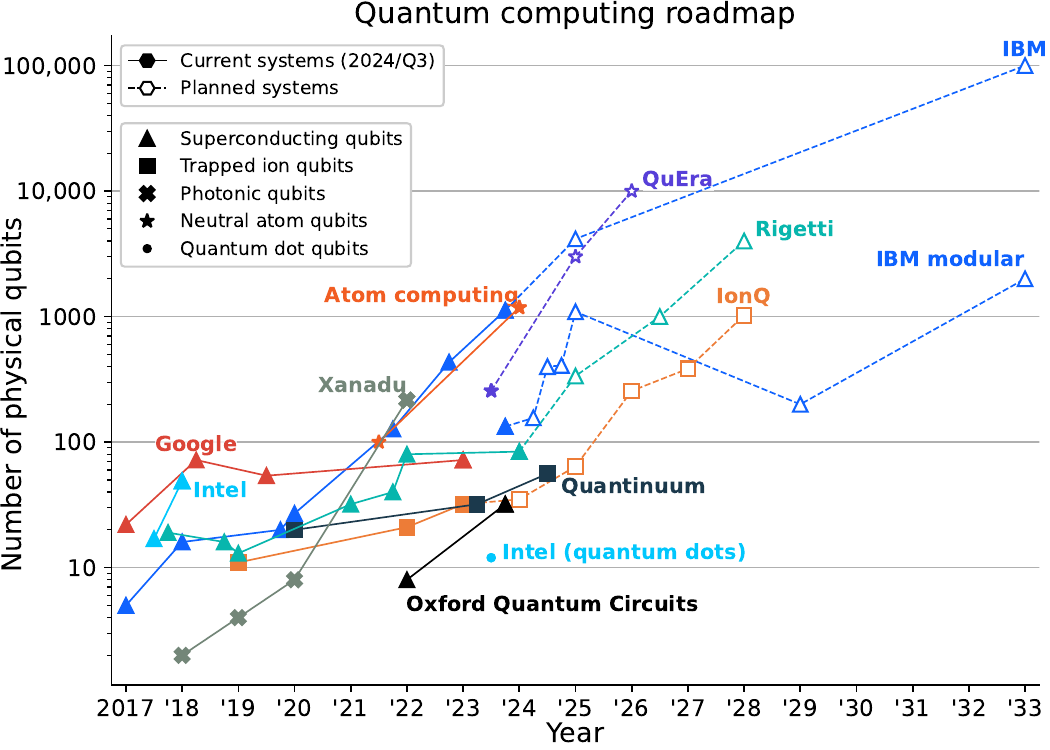}
    \caption{Quantum computing roadmap of the number of physical qubits by various manufacturers, inspired by~\cite{misc:hellstromroadmap}. Data can be found in \cref{tab:quantumroadmap}.
    }
    \label{fig:quantumroadmap}
\end{figure}

The first two quantities (the number of qubits and the error rate) are particularly important for the goal of achieving large-scale, reliable quantum computation. Increasingly, hardware manufacturers are providing technology roadmaps for the coming 5 to 10 years of development, with the number of qubits occupying a central role. In \cref{fig:quantumroadmap}, we illustrate the number of qubits planned by different manufacturers, using publicly available data. The first devices with more than one thousand qubits are already available, while within a decade there are commitments to building devices with up to \num{100000} qubits.  

\subsection{Quantum error correction}
The availability of quantum computers with hundreds of qubits, together with gate error rates around one percent, has begun to drive a wave of research in practical implementations of quantum error correction (QEC) codes~\cite{art:QuantinuumQEC, art:erhard2021entangling,art:krinner2022realizing,art:suppressingquantumerrors,art:postler2022demonstration, art:bluvstein2024logical,art:hetenyi2024creating}. A QEC code leverages multiple noisy physical qubits to encode a single \textit{logical qubit}, whose error rate is lower than that of a physical qubit. For a given QEC code, the number of physical qubits required to encode a logical qubit depends on the error rate of physical qubits, as well as the required error rate of the logical qubit~\cite{gottesman2009introduction, Devitt_2013, Roffe_2019}. Errors on logical qubits can be detected using a set of non-destructive `syndrome' measurements, while an accompanying decoder algorithm determines which operation(s) should be applied to correct an error. Quantum logic gates can be applied to a logical qubit by a suitable sequence of gates and measurements on the underlying physical qubits.

Quantum error correction will be essential to achieving quantum computational advantage for industry-sized problems. On the other hand, QEC carries a significant resource overhead, which must be carefully quantified to predict the scale at which a computational advantage may be obtained for a given use case. 

In this work, our resource estimation uses the surface code as the quantum error correction scheme. The surface code is a leading candidate for large-scale fault-tolerant quantum computation, and can be readily implemented on 2D planar devices such as superconducting qubit chips \cite{art:surfacecodes}.

\subsection{Quantum optimization for industrial shift scheduling}\label{QISS_section}

In this section we provide a brief overview of our industrial shift scheduling problem of interest, and the algorithm designed to solve it, the Quantum Industrial Shift Scheduling algorithm (QISS); full details may be found in~\cite{art:krol2024qiss}.

\begin{figure}[htbp]
    \centering
    \includegraphics[width=0.8\linewidth]{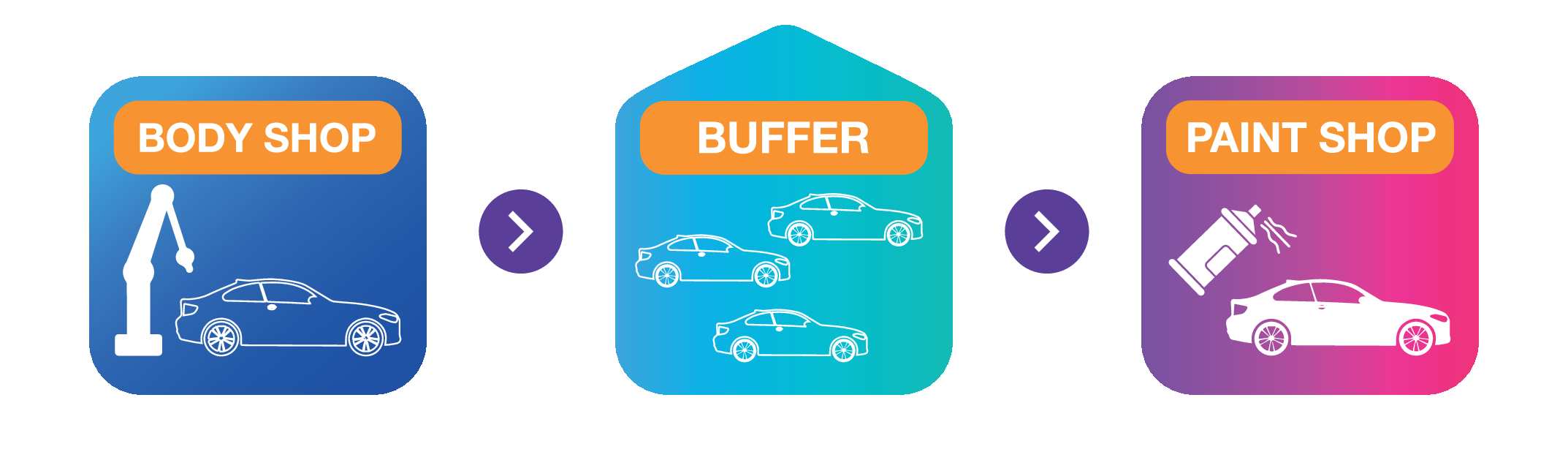}
    \caption{The structure of the simplified model for industrial shift scheduling, in which a body shop and a paint shop share a storage buffer~\cite{art:krol2024qiss}.}
    \label{fig:SSSM}
\end{figure}

We consider the simplified automotive production line shown in \cref{fig:SSSM}. This model consists of two \textit{shops}: a body shop and a paint shop, with a single shared storage buffer between them. 
To model the effect of labor regulations, each shop has a fixed set of four possible working hours (i.e.\ shifts) per day. Choosing to operate a shop for a given number of hours results in a corresponding production cost and vehicle output. The intermediate storage buffer cannot be filled beyond its maximum capacity, or emptied below its minimum capacity. The goal is to find a shift schedule for each shop such that an (annual) production volume target is met, up to some tolerance, with minimal operating costs.

QISS~\cite{art:krol2024qiss} uses Grover adaptive search to find the optimal solution with high probability, following a total of $O(\sqrt{N})$ applications of the corresponding Grover rotation operator. Here, $N$ is the total size of the solution space, where for $n$ days of factory operation we have $N = 16^n$; the constant factor of 16 represents the number of combinations of shifts for the two shops each day. In this work we use an adapted version of the QISS buffer constraint, which reduces the required number of ancilla qubits. This gives us a total qubit count of $6n+10+\log_2(19n)$ for scheduling $n$ days of factory operations~\cite{art:krol2024qiss}. This is explained in more detail in \cref{app:modifiedbufferconstraint}.

\section{Resource estimation for fault-tolerant quantum computing} \label{sec:resourceest}
Given a target quantum algorithm and an (assumed) quantum computer, the goal of resource estimation is to quantify the required size of the quantum computer and the runtime for complete execution of the computation. The target algorithm (QISS, in this work) is provided in the form of a quantum circuit using a high-level instruction set. The high-level instructions are convenient for the construction of algorithms, but do not take into account the need for quantum error correction, or the constraints imposed by the supported instruction set and the architecture of the quantum computer. More detail about resource estimation input is given in \cref{app:res_est_input}.

At the input stage the user must also specify the error budget, which can be determined independently based on the details of the target application. %
The error budget strongly influences the hardware requirements, because it determines the overhead that must be introduced to perform quantum error correction. The influence and division of our error budget is explained in more detail in \cref{appsec:errorbudgetdiv}.

For the most accurate estimation of the required resources, the quantum circuit is compiled into instructions that can be directly executed on the target hardware, in such a manner that the executable output circuit is fault-tolerant (i.e. robust to noise below a certain threshold). This compilation process consists of multiple steps, including: gate decomposition, qubit mapping and routing, resource scheduling and optimization, encoding into a quantum error correction scheme, and translation of gates into the device's native gateset~\cite{art:resourceestpaper, art:usingazureresest,art:openql}. However, at the present time many of the inputs to such a computation remain uncertain, given that quantum computers and the supporting software stack remain at an early stage of development. 
For quantum resource estimation, we instead make reasoned assumptions and approximations about the compilation procedure and the architecture of the quantum computer, allowing useful predictions to be obtained.   

\begin{figure}[htbp]
    \centering
    \includegraphics[width=\linewidth]{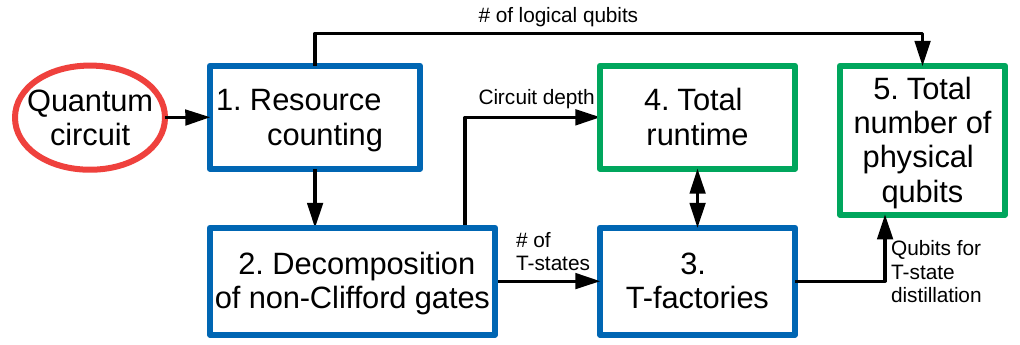}
    \caption{A simplified schematic overview of the process of resource estimation for fault-tolerant quantum computation. A more complete version of the diagram can be found in \cref{fig:res_est_diagram}.
    }
    \label{fig:simple_res_est_diagram}
\end{figure}

The process of quantum resource estimation can be summarized in the following five steps, shown schematically in \cref{fig:simple_res_est_diagram}.  A more elaborate explanation can be found in \cref{app:process_res_est}.

\begin{enumerate}[wide, labelindent=0pt]
    \item \textit{Logical resource counting: }
The first step in the process has two objectives: a) estimate the number of logical qubits required; and b) estimate the number of logical gates by reference to a standard instruction set. 

For a), an assumption is made about the underlying architecture of the quantum computer, and the additional logical ancilla qubits needed to facilitate interactions between logical qubits.
Meanwhile, b) uses the assumption that all operations in the target circuit will be implemented in practice via a standard instruction set of gates and/or measurements.

\item \textit{Decomposition of non-Clifford gates into Clifford and T-states: }
Unlike the set of Clifford gates, which in the surface code can be applied directly to logical qubits, non-Clifford gates must be implemented through a combination of Clifford gates and the use of \textit{magic} states~\cite{art:kitaev,art:universalqcwithidealclifford}, with the most commonly used magic state being the \textit{T-state}. 

To decompose the non-Clifford gates, a sequence of Clifford gates and T-states needs to be found that can approximate the non-Clifford gate to the required fidelity. This is achieved using a recursive algorithm whose runtime grows with the specified precision~\cite{art:ancillafreeSK,art:SKalg}.
For large circuits that require a very high fidelity, it becomes infeasible to decompose all non-Clifford gates and the number of T-states is instead based on existing results from the literature~\cite{art:resourceestpaper,art:GoogleBeyondQuadratic}.

\item \textit{Determination of the number and type of T-state factories: }
To produce a T-state, a physical T-gate must first be applied to a single physical qubit. The resulting quantum state is subsequently extended across multiple physical qubits to yield a logical qubit encoding the desired information~\cite{horsman_latticesurgery}. The logical qubit inherits the noise of the physical operation, and high-quality T-states need to be purified from multiple noisy T-states in \textit{T-state distillation factories}.
Different ways of implementing T-state factories have been proposed, each with differences in the required number of qubits and the runtime of the state distillation~\cite{art:resourceestpaper,art:universalqcwithidealclifford,art:GidneyFowler}.

\item \textit{Estimating runtime: }
The runtime of the algorithm is limited by one of two parallel processes: the circuit execution or the T-state distillation. 

The number of distillation factories can be chosen to minimize the number of qubits, by using only one T-state factory. If the goal is instead to minimize the runtime, as in this paper, the number of T-state distillation factories can be calculated so that they are able to produce all needed T-states in the same time as the circuit execution. 

\item \textit{Total number of physical qubits: }
The total number of physical qubits required to execute the algorithm depends on a number of factors. The most important are the required error rate of the logical qubits and the QEC scheme, which determine how many physical qubits are used to encode each logical qubit. 

The number of physical qubits for the circuit execution and the T-state distillation are calculated separately and then added together to find the total number of qubits required.
\end{enumerate}

\section{Near-term superconducting qubits}\label{sec:currentscens}
We base our quantum resource estimates on superconducting qubit technology, a choice that we make for several reasons. Firstly, multiple hardware developers have long-term roadmaps for superconducting qubit technologies, as shown in \cref{fig:quantumroadmap}. Secondly, previous quantum resource estimation results~\cite{art:resourceestpaper} have shown that superconducting qubits perform consistently well in terms of both the total computational runtime and the number of physical qubits, which we have also verified for our use-case (see \cref{app:predefinedscens}). Finally, published characteristics of superconducting qubits are available from several different sources, allowing us to extrapolate from current technologies to make our estimates. 

At the time of writing, no superconducting qubit quantum computer has surpassed the threshold for error correction with the surface code, in terms of all of the relevant physical error rates~\cite{art:surfacecodes,art:highthresholdqec,art:surfacecodeover1perc}. Therefore, the scenarios that we consider for resource estimation relate to future systems, where all figures of merit are below the threshold for QEC. Previous fault-tolerant resource estimation work has assumed all gates and operations to have identical error rates~\cite{art:resourceestpaper}.  We use two of these scenarios: the superconducting qubits with `flat' error rates of $10^{-3}$ and $10^{-4}$. Besides these, we also consider parameter sets that better reflect the profile of existing systems.

Specifically, we construct two scenarios based on the current state-of-the-art parameters reported in~\cite{art:suppressingquantumerrors}. The first scenario is obtained by decreasing these reference error rates by one order of magnitude; we refer to this as the `reduced error rate' (RER) scenario. The second scenario is based on a hybrid parameter set combining the figures of merit in~\cite{art:suppressingquantumerrors} and~\cite{art:transmonqubitreadoutfidelity}. Concretely, \cite{art:transmonqubitreadoutfidelity} introduces a qubit measurement technique with a shorter duration and higher fidelity than~\cite{art:suppressingquantumerrors}, as well as an improved qubit idle error rate that we calculate from the reported decay time of \qty{85.8}{\mu s}. The remainder of the parameters in this second scenario are drawn from~\cite{art:suppressingquantumerrors}; in particular, to ensure that the two-qubit error rate is below the surface code threshold, we use the value without crosstalk reported in that reference. We refer to this second parameter set as the `fast measurement' (FM) scenario. The two scenarios effectively allow us to separately probe the influence of improved error rates and faster measurements on the resources required. The parameter values for all scenarios are given in \cref{tab:paramsupercond}.

\begin{table}[tbph]
\caption{Parameters corresponding to the different scenarios for near-term superconducting qubits, of which results are plotted in \cref{fig:resultsQAI_e4,fig:code_distance_nearterm}. 'Error rate' is abbreviated with `ER', values in bold are those changed compared to the reference scenario~\cite{art:suppressingquantumerrors}.
} 
\label{tab:paramsupercond}
\centering
\setlength{\tabcolsep}{2.3pt}
\begin{tabular}{llllll}
\toprule
Name                 & \begin{tabular}[c]{@{}l@{}}Flat\\ $10^{\shortminus3}$\cite{art:resourceestpaper}\end{tabular} & \begin{tabular}[c]{@{}l@{}}Flat\\ $10^{\shortminus4}$\cite{art:resourceestpaper}\end{tabular} & \begin{tabular}[c]{@{}l@{}}Ref. \cite{art:suppressingquantumerrors}\end{tabular}    & \begin{tabular}[c]{@{}l@{}} RER\cite{art:suppressingquantumerrors}\end{tabular} & \begin{tabular}[c]{@{}l@{}}FM\cite{art:suppressingquantumerrors,art:transmonqubitreadoutfidelity}\end{tabular} \\ \midrule

1Q gate ER   & $10^{\shortminus3}$                                                                & $10^{\shortminus4}$                                                                  & $1.09 \times 10^{\shortminus3}$ & $1.09 \times 10^{\shortminus\bm{4}}$                                               & $1.09 \times 10^{\shortminus3}$                                               \\
2Q gate ER   & $10^{\shortminus3}$                                                                & $10^{\shortminus4}$                                                                & $6.05 \times 10^{\shortminus3}$ & $6.05 \times 10^{\shortminus\bm{4}}$                                               & $\bm{4.90}\times 10^{\shortminus\bm{3}}$                                      \\
Idle ER      & $10^{\shortminus3}$                                                                & $10^{\shortminus4}$                                                                & $2.46 \times 10^{\shortminus2}$ & $2.46 \times 10^{\shortminus\bm{3}}$                                               & $\bm{1.63}\times 10^{\shortminus\bm{3}}$                                      \\ 
1Q meas.\ ER & $10^{\shortminus3}$                                                                & $10^{\shortminus4}$                                                                & $1.96 \times 10^{\shortminus2}$ & $1.96 \times 10^{\shortminus\bm{3}}$                                               & $\bm{5.00}\times 10^{\shortminus\bm{3}}$                                         \\ \midrule
1Q meas.\ time       & \qty{100}{ns}                                                              & \qty{100}{ns}                                                              & \qty{500}{ns}  & \qty{500}{ns}                                                         & $\bm{140}\mskip\thinmuskip\textbf{ns}$                                       \\
1Q gate time         & \qty{50}{ns}                                                               & \qty{50}{ns}                                                               & \qty{25}{ns}   & \qty{25}{ns}                                                          & \qty{25}{ns}                                                 \\
2Q gate time         & \qty{50}{ns}                                                               & \qty{50}{ns}                                                               & \qty{34}{ns}   & \qty{34}{ns}                                                          & \qty{34}{ns}                                                 \\
\bottomrule
\end{tabular}
\end{table}
We begin by investigating the resource requirements as a function of the number of days of operation in our industrial shift scheduling problem, for a single application of the Grover rotation operator of the QISS algorithm. To interpret the results, we recall two of the basic space-time properties of the surface code. Firstly, for a surface code of distance $d$, $d$ rounds of syndrome measurements are performed to protect against measurement errors. As a result, the execution time for a logical cycle grows linearly with $d$. Secondly, the number of physical qubits required to construct a logical qubit grows proportional to $d^2$. 

\begin{figure}[!hbt]
    \centering
        \includegraphics[width=\linewidth]{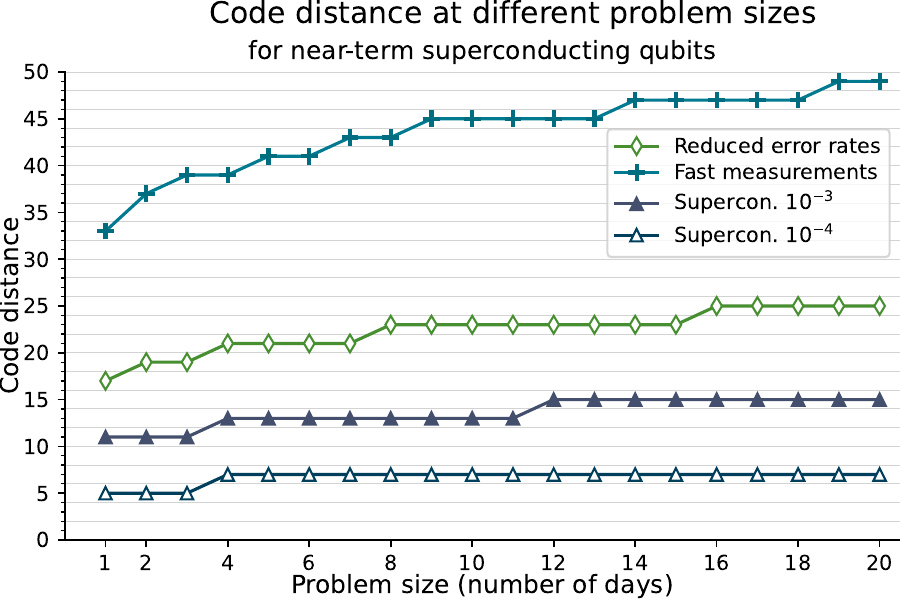}
    \caption{Code distance for the resource estimation of a single Grover rotation with the near-term qubit scenarios, with characteristics as outlined in \cref{tab:paramsupercond}. }
    \label{fig:code_distance_nearterm}
\end{figure}

\begin{figure}[!htb]
    \centering
    \begin{subfigure}[t]{\linewidth}
    \centering 
    \includegraphics[width=\linewidth]{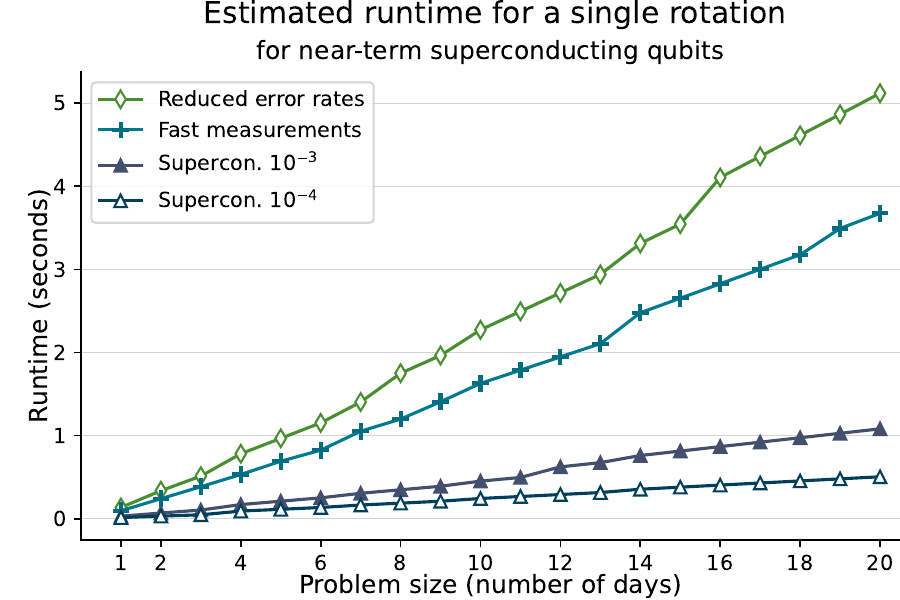}
    \caption{Estimated runtime for a single Grover rotation.}
    \label{fig:estimated_runtime_supercond}
    \end{subfigure}
    \hfill
    \begin{subfigure}[t]{\linewidth}
    \centering  
    \includegraphics[width=\linewidth]{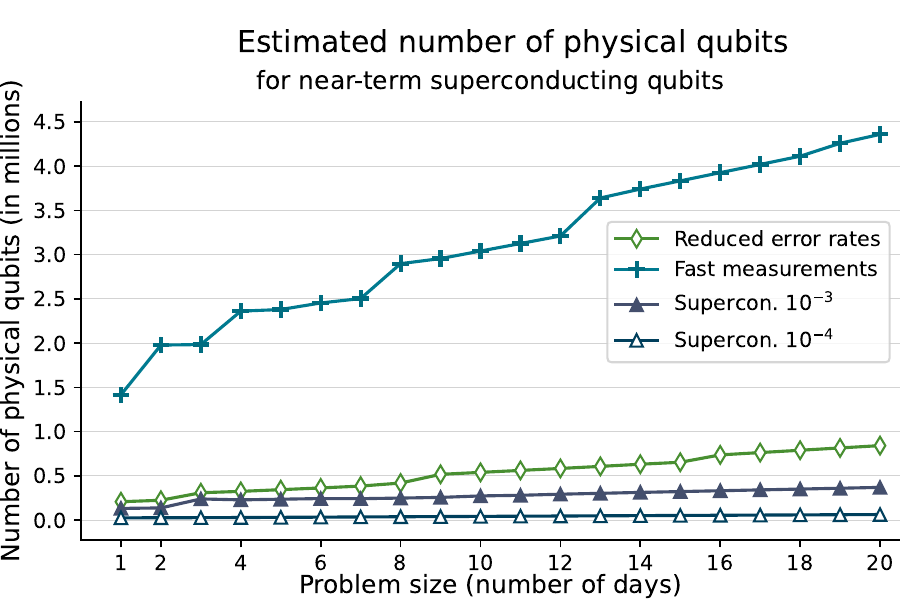}
    \caption{Estimated number of physical qubits for a single Grover rotation.}
    \label{fig:estimated_phys_supercond}
    \end{subfigure}
    \caption{Resource estimation results for the scenarios with characteristics given in \cref{tab:paramsupercond}.}\label{fig:resultsQAI_e4}
\end{figure}

The results for the different scenarios are shown in \cref{fig:resultsQAI_e4}. All computations were performed using the Microsoft Azure Quantum Resource Estimator~\cite{art:resourceestpaper}, which we have selected after comparing different resource estimation tools (see \cref{app:automated_res_tools}). The total runtime for each case has an approximately linear growth in the problem size, which can be partly explained by the linear increase of the number of logical operations in the Grover oracle~\cite{art:krol2024qiss}, which is the same for all scenarios. The total runtime is determined by multiplying the number of logical cycles by the execution time for a single cycle, which in turn is a function of the gate and measurement times and the code distance. 
For larger problem sizes, a larger code distance is generally required to meet the error budget. However, because the code distance is always rounded up to the nearest odd number, the required code distance increases in steps and the same distance is used for a range of problem sizes, as we show in \cref{fig:code_distance_nearterm}. 
This results in the piecewise linear growth of the runtime shown in \cref{fig:resultsQAI_e4}.

Comparing the results for the two flat error rate scenarios, the runtime for the $10^{-4}$ case is roughly half that of the $10^{-3}$ case, because the required code distance for the former is found to be around half that of the latter (7 vs 13 and 15, see \cref{fig:code_distance_nearterm}). Meanwhile, the runtime for the FM scenario grows more mildly than that of the RER scenario. For the particular parameters we have chosen, 
the longer measurement time of RER means that its logical cycle time is always larger. 

The total number of physical qubits required is computed as the sum of two contributions: 1) the number of physical qubits required by the circuit itself; and 2) the number of qubits required by the $T$-state factories. In turn, contribution 1) has two components: one that grows linearly in the number of days, resulting from the need to encode a larger number of shift choices and to check the corresponding optimization constraints, and another that grows quadratically in the required code distance, resulting from the encoding of logical qubits to physical qubits. The required code distance also has a dependency on the number of days, through the error rate required to meet the error budget, as can be seen in \cref{fig:code_distance_nearterm}. 
Meanwhile, contribution 2) depends on the required code distance and on the rate at which $T$ states are consumed. Because the number of operations in the Grover oracle increases linearly with the problem size~\cite{art:krol2024qiss},
the $T$-state consumption rate remains constant. Overall, we find that the number of physical qubits required in each scenario also follows an approximately linear growth in the number of days, as shown in \cref{fig:resultsQAI_e4}.

From these results, it is clear that industrial shift scheduling requires a number of qubits that (far) exceeds the technology that will be available in the near future (see \cref{fig:quantumroadmap}). Furthermore, running the QISS algorithm until the best solution is found will require ${\sim}\sqrt{N}$ Grover’s rotations~\cite{art:krol2024qiss}, and for scenarios that are feasible in the near-term (such as those considered in this section), the runtime of a single Grover rotation is already on the order of seconds.

\section{High-fidelity qubits} \label{sec:futurescens}
In this section, we turn to consider the resources required to run the QISS algorithm until completion, i.e. to execute the ${\sim}\sqrt{N}$ Grover rotations necessary to return the optimal solution with high probability. Our goal is to identify the characteristics of a system with high-fidelity qubits that would theoretically enable a quantum speedup for the industrial shift scheduling problem. We consider five scenarios with flat error rates up to $10^{-9}$, and a scenario with perfect qubits with an error rate of zero. 
Due to limitations of the resource estimator, we use a scaled error budget to obtain the resources required to execute the $\sqrt{N}$ iterations. This is explained in detail in \cref{app:scalingtosqrtN}.

The flat error rate scenarios investigated in the previous section had gate error rates of $10^{-3}$ and $10^{-4}$, gate execution times of \qty{50}{\ns}, and measurement times of \qty{100}{ns}~\cite{art:resourceestpaper}. We consider three additional high-fidelity scenarios, with error rates of $10^{-6}$, $10^{-8}$ and $10^{-9}$. We further assume that advances in qubit technology will lead to an improvement in all figures of merit simultaneously, taking the measurement time to be \qty{10}{\ns} for the $10^{-6}$ and $10^{-8}$ scenarios and \qty{1}{\ns} for the $10^{-9}$ scenario. The execution time for all gate types is half of the specified measurement time, i.e. \qty{5}{\ns} and \qty{0.5}{\ns}. 
Our assumptions, the construction and execution of the computations for these high-fidelity scenarios are discussed in more detail in \cref{app:futurequbitdetail}.

\begin{figure}[!hbt]
    \centering
    \begin{subfigure}[t]{\linewidth}
    \centering  
        \includegraphics[width=\linewidth]{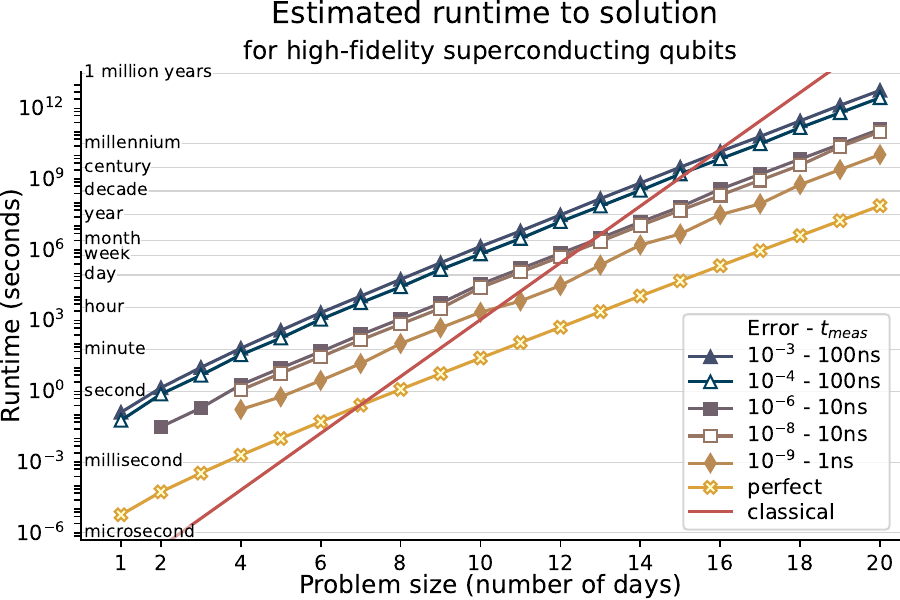}
        \caption{Estimated runtime for arriving at the best solution.}
        \label{fig:estimated_runtime_e3_e9}
    \end{subfigure}
    \begin{subfigure}[t]{\linewidth}
    \centering  
    \includegraphics[width=\linewidth]{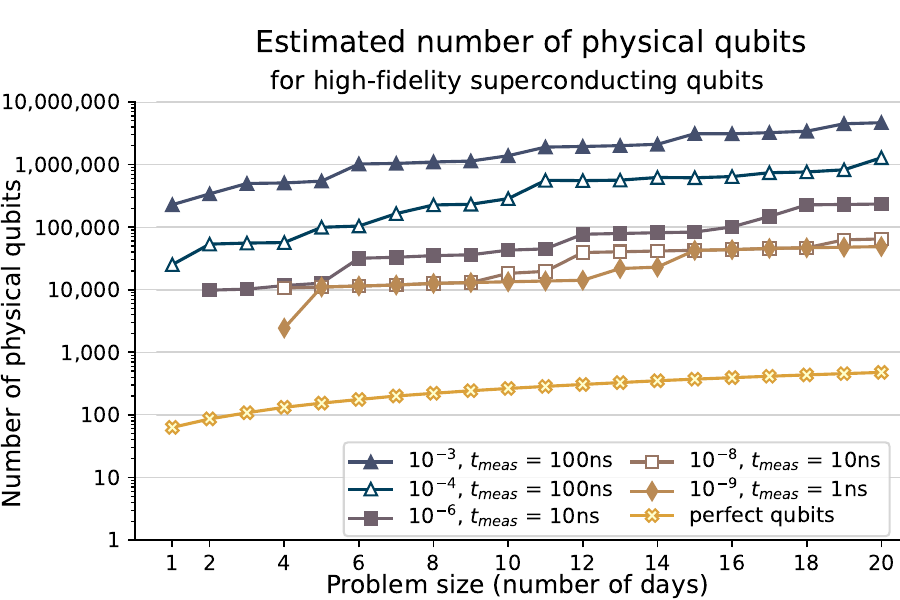}
    \caption{Estimated number of qubits for arriving at the best solution.}
    \label{fig:estimated_qubits_e3_e9}
    \end{subfigure}
    \caption{Resource estimation results for different scenarios based on high-fidelity superconducting qubits, with error rates from $10^{-3}$ to $10^{-9}$ and measurement times of \qty{100}{\ns} to \qty{1}{\ns}. The execution time for all gate types is half of the specified measurement time. Full characteristics can be found in \cref{tab:paramse3_e9}.
    }
    \label{fig:results_e3_e9}
\end{figure}

The results of the resource estimation with the high-fidelity scenarios are shown in \cref{fig:results_e3_e9}. We compare the runtime with a classical brute-force search, assuming a set of parallel processes with a combined completion frequency of \qty{1}{GHz}, meaning that $10^9$ guesses at the best solution are completed every second. For error rates of $10^{-3}$ and $10^{-4}$, the number of physical qubits is of the order of a few million for problem sizes up to 20 days. However, the quantum algorithm does not exhibit a speedup over classical brute-force search until the complete computation takes a over a century. For error rates between $10^{-6}$ and $10^{-8}$ with a measurement time of \qty{10}{\ns}, the quantum algorithm becomes faster than the classical brute-force search around a problem size of 12 days, with execution times on the order of weeks. For a very low physical error rate of $10^{-9}$, the quantum algorithm achieves a speedup for scheduling 11 days, taking a little over 2 hours compared to almost 5 hours using the classical approach. A quantum computer with completely error-free qubits would enable a speedup over classical brute-force search for problem sizes larger than 7 days. However, even in this case the total runtime would be of the order of months for scheduling 20 days of operations.

\begin{figure}[!hbt]
    \centering
        \includegraphics[width=\linewidth]{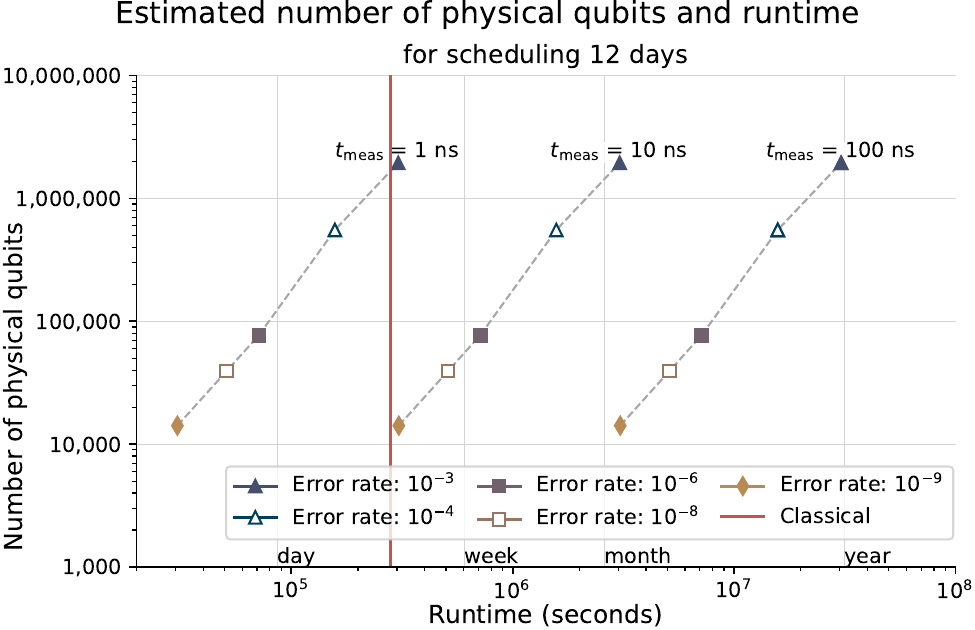}
    \caption{Resource estimator results for scheduling twelve days for superconducting qubits error rates from $10^{-3}$ to $10^{-9}$ and measurement times of \qty{1}{\ns}, \qty{10}{\ns} and \qty{100}{\ns}. The execution time for all gate types is half of the specified measurement time.  Full characteristics can be found in \cref{tab:paramse3_e9}. 
    } 
    \label{fig:physical_qs_v_runtime12}
\end{figure}

To further investigate the influence of error rates and measurement execution times, we focus on the particular case of scheduling for 12 days of operation. We consider different combinations of the two system figures of merit, and plot the runtime and the number of physical qubits in \cref{fig:physical_qs_v_runtime12}. When a measurement takes only \qty{1}{\ns}, all of the tested scenarios are faster than the brute-force approach for this problem size, except the one with an error rate of $10^{-3}$. When measurements take \qty{10}{\ns} or more, all scenarios are slower than the classical approach. These results clearly show that the error rate has less influence on the overall runtime than the measurement execution time. For example, decreasing the error rate from $10^{-3}$ to $10^{-9}$ gives a speedup of approximately an order of magnitude, whereas a reduction in measurement time translates directly into the same reduction in total runtime. 

\section{Conclusion and outlook} \label{sec:conclusion} 
This work was motivated by the question of whether a speedup could realistically be achieved for industrial shift scheduling using the QISS algorithm. We have compared a number of scenarios with different system parameters, covering quantum computers that could feasibly be available in the near or medium term, as well as hypothetical high-fidelity scenarios that would require significant technological improvements. We find conclusively that the near-term scenarios will not deliver a quantum speedup, while even the high-fidelity scenarios would require execution time of measurement operations to be below \qty{10}{ns}. This is in line with existing work investigating the plausibility of quantum speedups using Grover-based algorithms, which emphasized that the slow execution time of quantum operations compared to classical ones is a major obstacle for such approaches~\cite{art:TroyerBeyondQuadratic, art:GoogleBeyondQuadratic}.

Besides hardware improvements to reduce gate and measurement times, there are two further areas where future work could significantly change the prospects for a quantum speedup for large industrial optimization problems. Firstly, novel quantum error correction schemes could result in a faster logical clock cycle, which could occur if, for example, the required code distance is smaller. In turn, this could be achieved for QEC codes with a higher threshold than the surface code, although we emphasize that the cycle time for a QEC code is determined by many other factors besides its threshold. Secondly, advances in quantum algorithms could lead to approaches that achieve beyond-quadratic speedups, potentially significantly reducing the crossover point at which a practical quantum speedup can be realized.

For future work, one can consider interesting experimentation by executing small instances of the QISS algorithm on quantum computers that will be available within the next decade. For example, the results of \cref{fig:results_e3_e9} suggest that in the $10^{-3}$ and $10^{-4}$ flat error rate scenarios (which may be considered realistic with hardware improvements) one could execute the algorithm on problem sizes of around 7 to 9 days, with a runtime in the order of hours or days. It may then be of interest to systematically study the impact of executing fewer than the maximum number of Grover iterations, in order to determine the trade-off in runtime and solution quality. With fewer iterations the probability of obtaining the optimal solution is reduced, however in practice one may be interested instead in obtaining a good approximate solution. In this context, one could investigate the probability of obtaining a solution satisfying some acceptable approximation ratio, as a function of the number of iterations and the number of trials, where independent trials may also be parallelized over multiple quantum processors. It may then be of interest to compare the performance of this heuristic quantum approach to that of classical heuristics.

\printbibliography
\end{refsection}

\begin{refsection}

\onecolumn
\appendices 
\renewcommand\thesubsection{\thesection.\Roman{subsection}}
\renewcommand\thesubsectiondis{\thesection.\Roman{subsection}}

\section{Modified buffer constraint} \label[appendix]{app:modifiedbufferconstraint}
Because of (hardcoded) limitations in the quantum simulator and the resource estimation tools that we used, we need to reduce the number of qubits used by the QISS algorithm~\cite{art:krol2024qiss}. With the reduction in qubits, we can check the correctness of our algorithm for larger problem sizes than would otherwise be possible because of these limitations (see also \cref{app:scalingtosqrtN}). 

To reduce the number of qubits required in the QISS algorithm, we introduce an additional constraint that disqualifies any solution that results in a negative number of units in the buffer. In practice, this means that shop 2, which takes units out of the buffer, is not allowed to be idle (i.e. assigned to work, but with no stock to work on). 

In~\cite{art:krol2024qiss}, the number of units in the buffer is set to zero if it would otherwise hold a negative number, using the $MAX(0,\tilde{B})$ operator. Because the oracle needs to be a reversible computation, the $MAX(0,\tilde{B})$ operator requires many ancilla qubits to store this negative buffer value. But if the number of units in the buffer exceeds the maximum value of $B_{max}$, the solution is disqualified through use of a single condition qubit per day of scheduling. 

We can use the same construction for a minimum number of items in the buffer. This means that the $MAX(0,\tilde{B})$ operator and all its ancilla qubits can be replaced by a single condition qubit and corresponding constraint check.
The modified circuit, shown in \cref{fig:circuitfornomaxb}. 
This modification reduces the need for the $(6n-1)$ ancilla qubits for scheduling $n$ days, while adding one additional condition qubit per day. This reduces the total qubit count by approximately $45\%$, from $11n+9+\log_2(19n)$ to $6n+10+\log_2(19n)$.

\begin{circuit}[hbt]
\[
\begin{array}{c}
\Qcircuit @C=0.3em @R=0.5em {
&\lstick{\ket{q_{S1}}} & \qw & \ctrlxo{2} & \qw &\qw & \qw& \qw& \qw & \qw & \qw & \qw & \qw & \qw & \qw \\
&\lstick{\ket{q_{S2}}} & \qw &\qw & \ctrlxo{1} & \qw & \qw& \qw& \qw & \qw & \qw & \qw & \qw & \qw & \qw \\
&\lstick{\ket{q_{b}}} & \gate{B_\text{init}} & \gate{S1} & \gate{S2} & \gate{QFT^\dagger} &\ctrl{2} &\gate{QFT}& \gate{\shortminus(B_\text{max}+1)} & \gate{QFT^\dagger} & \ctrl{3} & \gate{QFT} &  \gate{(B_\text{max}+1)} & \qw  & \qw\\
&\lstick{\color[rgb]{0.4,0.4,0.4}\ket{a}}  & \qwdash & \qwdash & \qwdash& \qwdash& \qwdash& \qwdash & \qwdash & \qwdash& \qwdash & \qwdash & \qwdash & \qwdash & \qwdash\\
&\lstick{\ket{c_{1,\text{min}}}}  & \qw & \qw & \qw& \gate{X} & \targ & \qw & \qw & \qw & \qw & \qw &\rstick{\ket{1} \textrm{ if }B_\text{out} \geq 0 }\qw\\
&\lstick{\ket{c_{1,\text{max}}}}  & \qw & \qw& \qw & \qw & \qw & \qw & \qw& \qw& \targ & \qw & \rstick{\ket{1} \textrm{ if }B_\text{out} \leq B_\text{max}} \qw\\
}
\end{array}
\]
\caption{Circuit for the evaluation of the modified minimum buffer capacity condition $B_\text{out} \geq 0$, with gates as defined in~\cite{art:krol2024qiss}. The CNOTs are controlled by the most significant bit (MSB) of the register $\ket{q_b}$. The buffer register is returned to its original value $B_\text{out}$ through a quantum Fourier transform and by adding the value of $B_\text{max}+1$. The original formulation required the use of a register of ancilla qubits $\ket{a}$ to implement the $MAX(0, \tilde{B})$ operator. With this modification, these ancilla qubits are not required and can be omitted, 
at the cost of adding one additional condition qubit ($\ket{c_{1,\text{min}}}$) per day of scheduling.
}
    \label{fig:circuitfornomaxb}
\end{circuit}

\section{Resource estimation input} \label[appendix]{app:res_est_input}
Besides the target quantum circuit encoding the computation to be performed, the key inputs to resource estimation include the
characteristics of the target quantum computer to be used, the error budget for the computation, and the quantum error correction code. 

\subsection{Characteristics of target device}

The key properties of the target device that can be configured in the resource estimation tools presented in \cref{app:automated_res_tools} include the standard error rates due to gate operations, measurements, and qubit idling. Additionally, one must specify an error rate for physical T-gates, which influences the resources required for T-state distillation. When not otherwise specified, we assume T-gate error rates are the same as 1-qubit gate error rates, and that the idle error rate is the same as the measurement error rate.  

\subsection{Error budget}
The error budget is the allowed error rate after execution of the whole circuit, and represents the fraction of times the circuit is allowed to fail (i.e. to return an incorrect answer). It is the sum of three components: the error budget for logical qubits, the budget for rotation gate synthesis, and the budget for T-state distillation. In our computations, the total error budget $\epsilon$ is divided equally among the three components, such that each has a value of $\epsilon_{\mathrm{log}}=\epsilon_{\mathrm{dis}}=\epsilon_{\mathrm{rot}}= \epsilon/3$~\cite{art:resourceestpaper}. We illustrate the influence of these three quantities in the resource estimation process in \cref{fig:res_est_diagram}, and provide more discussion in \cref{app:errorbudget}.

For Grover-based algorithms such as QISS, it is relatively easy to verify whether the solution obtained satisfies the desired search criteria. In our case, we seek a solution with an objective function value that is lower than that of any previously found solution.
This means that we have a relatively high tolerance for errors in our algorithm. Therefore, we choose an error budget of $\epsilon = 0.25$ for our application.

Increasing the error budget increases the average time to solution, and decreases the chance that the optimal solution has been found after $\sqrt{N}$ application of the Grover rotation operator. We have accounted for this by setting the number of rotations to $\sqrt{N}$ rather than $\frac{\pi}{4}\sqrt{N}\approx 0.79\sqrt{N}$ for our algorithm. 

\subsection{Quantum error correction code} \label{subsec:qec}
A number of parameters of the chosen QEC code influence the corresponding resource requirements, including the code threshold and the crossing pre-factor. Ultimately, these determine the code distance required to satisfy the error budget for the computation. In turn, the code distance strongly influences the encoding overhead (i.e. the ratio of physical to logical qubits), and the logical cycle time (i.e. the time required to perform operations such as error detection, correction, and logical gates).

In all of our experiments we focus on the surface code, which is a leading candidate for large-scale fault-tolerant quantum computing ~\cite{art:kitaevfaulttolerantquantumcomputing, art:surfacecodes}. We make use of the in-built surface code models in the Azure Quantum Resource Estimator (AQRE), which are described in detail in~\cite{misc:resestdocumentation}.

\section{Process of resource estimation}
\begin{figure*}[ht]
    \centering
    \includegraphics[width=\linewidth]{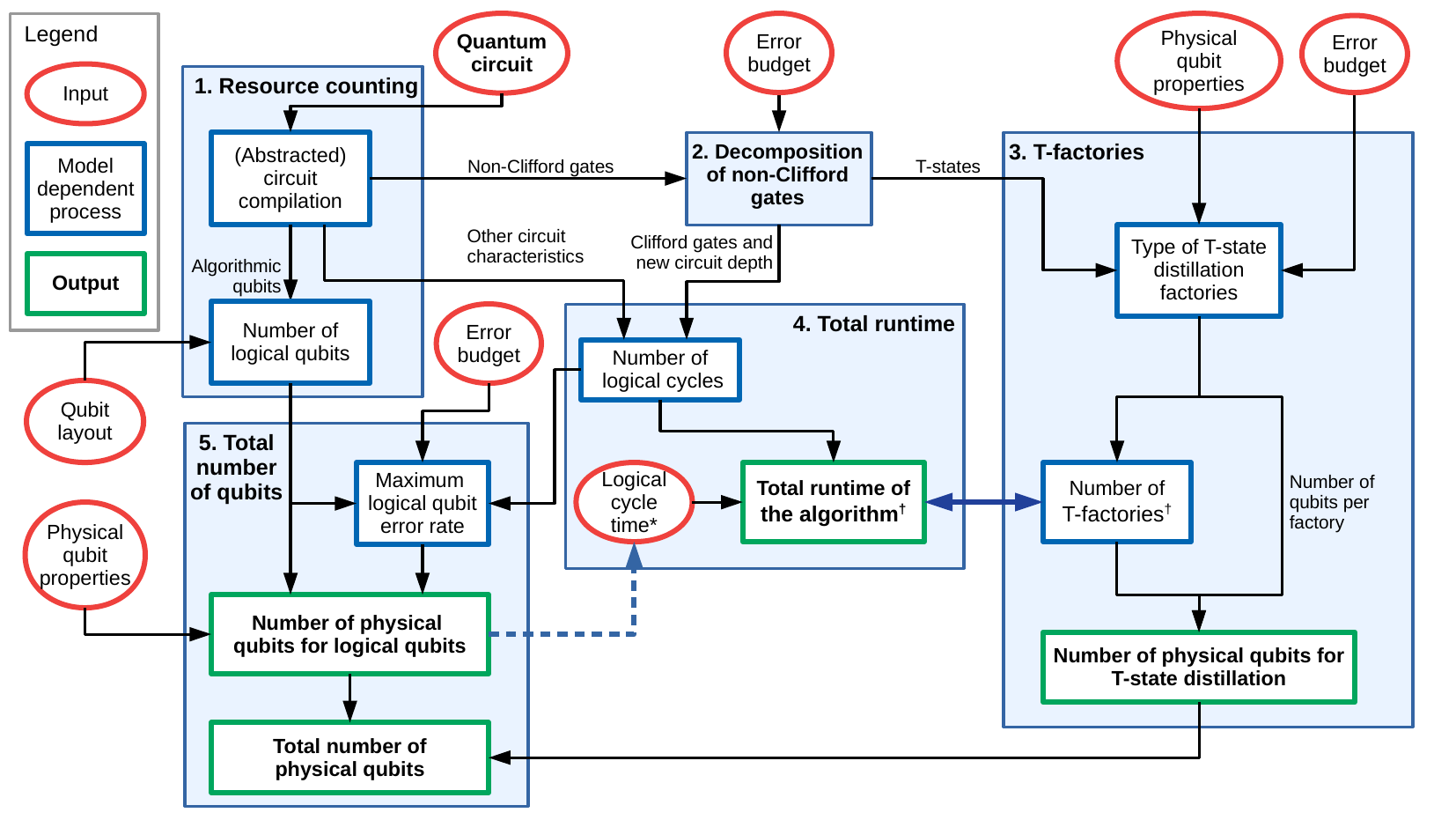}
    \caption{Schematic overview of the process of resource estimation as used in the Microsoft Azure Quantum Resource Estimator~\cite{art:resourceestpaper}, Qualtran~\cite{misc:qualtran} and Bench-Q~\cite{misc:benchq}. Red outlined circles show the (user) input, blue outlines mark the model dependent processes and green outlines show the output of the estimation.\\{\footnotesize * Logical cycle time can be an input value or can be calculated from the (input) gate characteristics, the code distance required to achieve the logical qubit error rate and the characteristics of the error correction code.\\
    $\dagger$ The algorithm runtime can be calculated by which takes the most time: the logical circuit execution or the T-state distillation. Alternatively, the number of distillation factories can be chosen so that the T-state distillation does not increase the total runtime, but takes less than or as much time as the logical circuit execution.}} \label{fig:res_est_diagram}
\end{figure*}

\label[appendix]{app:process_res_est}
A schematic overview of the process of quantum resource estimation is shown in \cref{fig:res_est_diagram}. In this appendix, we first explain in detail the different steps of the process, and then give an example for a problem size of ten days.

\subsection{Resource estimation in five steps}
\noindent The procedure of quantum resource estimation can be understood as consisting of five main steps:

\begin{enumerate}
    \item Counting of the logical resources required by the input circuit (i.e. the circuit that encodes the target application); 
    \item (Abstracted) decomposition of all non-Clifford gates in the input circuit into sequences of Clifford gates and T-states;
    \item Determination of the type and number of T-state factories required to supply the required T-states within the runtime of the computation;
    \item Calculation of the total estimated runtime for executing the complete circuit;
    \item Calculation of the total number of physical qubits required.
\end{enumerate}%
Note that this division of steps is a simplification of the full process of resource estimation. Other approaches may, for example, use a different subdivision of the process, or use similar steps in a different order. 

\subsubsection{Logical resource counting}
The first step in the process is the circuit compilation, using the assumption that all operations in the target circuit will be implemented in practice via a standard instruction set of gates and/or measurements. For example, an arbitrary rotation around either the $Y$ or $X$ axis is decomposed into a product of Hadamard-type gates and a rotation around the $Z$ axis.

Additional logical ancilla qubits are needed to facilitate interactions between logical qubits. How many ancilla qubits are needed depends on the topology of the target device, and for the most accurate estimator results, a full mapping of the input circuit onto the target qubit layout is required. If the topology is not known or mapping is not feasible, assumptions about the qubit layout and the overhead introduced by mapping can be used instead. 

For example, one may assume a 2D planar device with nearest-neighbor connectivity, with alternating rows of logical algorithm qubits (i.e. those being used explicitly for computation) and logical ancilla qubits~\cite{art:resourceestpaper}.

\subsubsection{Decomposition into Clifford and T-states}
 
Unlike the set of Clifford gates, which in the surface code can be applied directly to logical qubits using methods of lattice surgery, non-Clifford gates must be implemented through a combination of Clifford gates and the use of \textit{magic states}~\cite{art:kitaev,art:universalqcwithidealclifford}, with the most commonly used magic state being the \textit{T-state}. 

To decompose non-Clifford gates, a sequence of Clifford gates and T-states needs to be found that can approximate the non-Clifford gate to the required fidelity. For Toffoli gates, a standard construction from the literature may be used~\cite{art:Toffoli_FourTs}. For other non-Clifford operations such as arbitrary rotations around the $Z$ axis, a decomposition is obtained using a recursive algorithm whose runtime grows with the specified precision~\cite{art:ancillafreeSK,art:SKalg}. 
For large circuits with stringent precision requirements, it becomes infeasible or impractical to decompose all non-Clifford gates. 
Instead, in resource estimation the error budget can be combined with existing results from the literature to calculate an average number of T-states required for the decomposition of each non-Clifford gate. The total number of required T-states can then easily be calculated by multiplying this average and the number of non-Clifford gates obtained in the resource counting step~\cite{art:resourceestpaper}.

\subsubsection{Determination of the number and type of T-state factories}

To implement the quantum circuit within the specified error budget, high-fidelity T-states are prepared using \textit{T-state factories}.

Since state injection begins with a non-error-corrected physical T-gate, the resulting logical state inherits the noise of the physical operation. Subsequently, high-quality T-states, which have a fidelity consistent with the target accuracy of the computation, are purified from multiple imperfect, noisy T-states using sophisticated quantum circuits known as \textit{T-state distillation factories}. Different ways of implementing T-state factories have been proposed, with the type of factory chosen influencing the required number of qubits and the runtime of each factory~\cite{art:resourceestpaper,art:universalqcwithidealclifford,art:GidneyFowler}.

\subsubsection{Estimating runtime}
The algorithm's runtime is limited by one of two parallel processes: the circuit's execution and the T-state distillation.
The circuit execution time depends on the degree of parallel gate execution that is assumed possible, the specific details of the quantum error correction code used, and the execution time of underlying physical operations such as two-qubit gates, measurements, and qubit resets.

The time needed for T-state distillation also depends on these factors. However, the number of distillation factories can be chosen to either minimize the number of qubits (by using only one T-state factory) or to minimize the runtime by finding the number of factories able to produce all needed T-states at the same time as the circuit execution. A common assumption for resource estimation is that the rate of T-state consumption is constant. In other words, we assume that the computation requires an equal number of T-states at any moment during its execution. One may relax this assumption and incorporate circuit-specific information on the T-state requirements during various stages of the computation, at the expense of introducing further modeling and computational complexity to the resource estimation procedure. 

\subsubsection{Total number of physical qubits}
The total number of physical qubits required to execute the computation depends on a number of factors. Primarily, any quantum error correction scheme has an intrinsic overhead, where multiple physical qubits are used to encode a single logical qubit. The resource footprint of this encoding overhead depends strongly on the error correction code being used, the error budget of the computation, and the predominant error rate of the physical qubits. 

Secondly, the algorithm design and compilation process can have a significant influence the required number of qubits. For example, the need for logical qubit routing can be reduced by designing the algorithm in such a way that each qubit only interacts with its nearest neighbors. Using more sophisticated mapping algorithms can reduce the number of physical qubits required for mapping, at the cost of increasing the runtime of the compilation process~\cite{art:optofqcfornearest}.
The device topology also influences the number of physical qubits. For architectures with high connectivity, the need for physical qubit routing may be reduced compared to e.g. a device with only one-dimensional (linear) qubit connectivity.

For resource estimation, we use assumptions about the influence of device connectivity and how it is used, instead of running the computationally expensive task of full mapping and scheduling of the algorithm.

The number of physical qubits for the circuit execution and the T-state distillation can be calculated separately and then added together to find the total number of qubits required.

\subsection{Analytical example of resource estimation} \label[appendix]{app:analyticalexample}
\begin{table*}[ht]
\caption{Input values, parameters and resource counting results for the analytical resource estimation example.} \label{tab:exampleinputvals}
\centering
\begin{tabular}{lll|lll}
\toprule
\textbf{Symbol}   & \textbf{Description}                                                     & \textbf{Value} & \textbf{Symbol} & \textbf{Description}                  & \textbf{Value} \\ \midrule
$n_\text{days}$   & Number of days to schedule                         & $10$           & $Q_\text{alg}$  & Number of algorithmic qubits                                             & 116            \\
$\epsilon$        & Error budget                                       & $0.001$        & $M_\text{meas}$ & Number of measurement operations                                         & 5858           \\
$p$               & Single qubit measurement error rate                & $0.001$        & $M_\text{R}$    & Number of single-qubit rotation gates                                    & 7938           \\
$p_T$             & T-state error rate                                 & $0.001$        & $M_\text{T}$    & Number of T-gates                                                        & 912            \\
$t_\text{2Qgate}$ & Two-qubit gate time                                & \qty{50}{\ns}  & $M_\text{Tof}$  & Number of Toffoli gates                                                  & 5820           \\
$t_\text{meas}$   & Measurement time                                   & \qty{100}{\ns} & $D_\text{R}$    & Rotation depth                                                           & 3653           \\
\midrule
$a$               & Surface code parameter~\cite{art:resourceestpaper} & $0.03$         & $A$             & Value for determining the number of T-states~\cite{art:resourceestpaper,art:shorterqcircuits} & $0.53$         \\
$p^*$             & Surface code parameter~\cite{art:resourceestpaper} & $0.01$         & $B$             & Value for determining the number of T-states~\cite{art:resourceestpaper,art:shorterqcircuits} & $5.3$     \\    
\bottomrule       
\end{tabular}
\end{table*}

To get some insight into how the resource estimates are calculated, we explicitly show the calculations going into the estimations for our industrial shift scheduling algorithm for a fixed problem size of ten days, and one application of the corresponding Grover oracle. This calculation follows the steps shown in \cref{fig:res_est_diagram}. The relevant input values for the example are listed in \cref{tab:exampleinputvals}, the intermediate and output values can be found in \cref{tab:outputsanalyticalexample}. 

The algorithm is written as a Q\# program and compiled to QIR using the standard Q\# compiler built into the AQRE. We outline the steps and calculations behind the resource estimator using this example, for the ``qubit\_gate\_ns\_e3" parameter set (i.e. the `Flat $10^{-3}$' scenario of \cref{tab:paramsupercond}). We set the error budget to be $\epsilon = 0.001$, which we divide into three equal parts into error budgets for the logical circuit execution $\epsilon_\text{log}$, the error budget for the approximation of the synthesis of the rotation gates $\epsilon_\text{syn}$ and the error budget for the T-state distillation $\epsilon_\text{dis}$:
\begin{align}
    \epsilon = \epsilon_\text{log} + \epsilon_\text{syn} + \epsilon_\text{dis}~,~ \epsilon_\text{log} = \epsilon_\text{syn} = \epsilon_\text{dis} = \frac{\epsilon}{3} = 0.000\overline{3}\label{eq:eb_divide}    
\end{align}

The number of logical qubits and gates were obtained using the logical counts as output by the AQRE. To improve algorithm performance the AQRE uses a compilation scheme that delegates the expensive non-Clifford rotation operations to ancilla qubits. To implement a rotation gate such as $R_z(\theta)$ to a qubit, the qubit is first entangled with an ancilla by a joint Pauli measurement, after which a series of Clifford and T-states is applied to the ancilla to apply the rotation $\theta$. By measuring the ancilla, the phase $\theta$ is kicked back to the algorithm qubit. By using this compilation technique, multiple rotations can be synthesized in parallel~\cite{art:resourceestpaper}. The ancilla qubits and measurement operations this technique requires are counted among the values listed in~\cref{tab:exampleinputvals}. 

For a problem size of ten days, this technique adds $38$ ancilla qubits to the $78$ qubits needed for the algorithm execution~\cite{art:krol2024qiss}, bringing the total to $116$ algorithmic qubits ($Q_\text{alg}$). The additional measurement operations account for most of the counted measurements operations: of the $M_\text{meas} = 5858$ measurements, $4 \cdot n_\text{days} = 40$ are for measuring the state of the shop qubits.

The AQRE determines the number of logical qubits $Q$, the minimum number of logical cycles $C_\text{min}$, and the required number $M$ of T-states as follows~\cite{art:resourceestpaper}:

\begin{align}
    Q &= 2 Q_\text{alg} + \left\lceil \sqrt{8 Q_\text{alg}} \right\rceil + 1 \label{eq:D1} \\
    C_\text{min} &= (M_\text{meas} + M_\text{R} + M_\text{T}) + \left\lceil A \log_2\left(\frac{M_\text{R}}{\epsilon_\text{syn}}\right) + B \right\rceil D_\text{R} + 3 M_\text{Tof} \label{eq:D3} \\
    M &= \left\lceil A \log_2\left(\frac{M_\text{R}}{\epsilon_\text{syn}}\right) + B \right\rceil M_\text{R} + 4 M_\text{Tof} + M_\text{T} \label{eq:D4}
\end{align}%
From \cref{eq:D1}, we can calculate that this circuit will require $Q=264$ logical qubits. Using \cref{eq:D3,eq:D4}, with the values listed in \cref{tab:exampleinputvals}, we calculate that the circuit will consume $M = \num{175014}$ T-states and take a minimum of $C_\text{min} = \num{101575}$ logical time steps. 
We will assume that the total number of logical time steps will be equal to the minimum required number for the logical circuit execution  (i.e. the T-state distillation will not increase the total runtime of the algorithm). 
That means that we can use the number of logical cycles and the number of logical qubits $Q$ to calculate the (maximum) logical qubit error rate $P$ as:
\begin{align}
     Q\cdot C_\text{min} \cdot P = \epsilon_{\log} \leq \epsilon /3 \label{eq:QCP}
\end{align}%
This gives us a maximum logical qubit error rate of $P = 1.24\cdot 10^{-11}$. 
Using \cref{eq:E2}, we find that we need a (minimum) code distance of $d=19$ to achieve the logical error rate. 
The required number of physical qubits per logical qubit is thus $n(d) = 722$ (\cref{eq:nd}), and the number of physical qubits required for encoding all of the logical qubits $q_\text{log} = Q \cdot n(d) = \num{190608}$.

\begin{align}
    d &= \left\lceil \frac{2 \log(a/P)}{\log(p^*/p)} -1 \right\rceil_{odd}    \label{eq:E2} \\
    n(d) &= 2d^2 \label{eq:nd}
\end{align}%
The logical cycle time for the surface code is calculated using \cref{eq:taud}, from the two-qubit gate time $t_\text{2Qgate}$ and the measurement time $t_\text{meas}$: $\tau(d) = (4\cdot \qty{50}{\ns}+2\cdot \qty{100}{\ns})\cdot 19 = \qty{7600}{\ns}$. This gives a total algorithm runtime of $t = \qty{7600}{\ns}\cdot \num{101575}= \qty{0.772}{\s}$ (\cref{eq:totalruntime}).

\begin{align}
    \tau(d) &= (4 \cdot t_\text{2Qgate}+2 \cdot t_\text{meas}) \cdot d \label{eq:taud} \\
    t &= \tau(d) \cdot C_\text{min} \label{eq:totalruntime}
\end{align}

\begin{table*}[bth]
\caption{Intermediate and output values for the analytical resource estimation example.} \label{tab:outputsanalyticalexample}
\centering
\begin{tabular}{lll|lll}
\toprule
\textbf{Symbol}             & \textbf{Description}                          & \textbf{Value}       & \textbf{Symbol}                                           & \textbf{Description}                          & \textbf{Value}      \\ \midrule
$\epsilon_\text{syn}$ & Error budget for rotation gate approximations & $0.000\overline{3}$  & $\epsilon_\text{dis}$                      & Error budget for T-state distillation         & $0.000\overline{3}$ \\
$\epsilon_\text{log}$ & Error budget for logical circuit execution    & $0.000\overline{3}$  & $M$                                        & Number of T-states                            & $\num{175014}$      \\
$Q$                   & Number of logical qubits                      & \multicolumn{1}{l|}{$264$}                & $F$                                        & Number of T-state distillation factories      & 23                  \\
$C_\text{min}$        & Minimum number of logical cycles              & \multicolumn{1}{l|}{$\num{101575}$}       & $\mathcal{D}_1$, $\mathcal{D}_\text{last}$ & Code distance for T-state distillation rounds & 5, 17               \\
$P$                   & Logical Clifford error rate                   & \multicolumn{1}{l|}{$1.24\cdot 10^{-11}$} & $P_T(\mathcal{D})$                         & Logical T-state error rate                    & $1.90\cdot 10^{-9}$ \\
$d$                   & Code distance                                 & \multicolumn{1}{l|}{$19$}                 & $M(\mathcal{D})$                           & Number of T-states produced per distillation                & $1$                 \\
$n(d)$                & Number of physical qubits per logical qubit   & \multicolumn{1}{l|}{$722$}                & $n(\mathcal{D})$                           & Number of qubits per factory                  & \num18000           \\
$\tau(d)$             & Logical cycle time                            & \multicolumn{1}{l|}{\qty{7600}{\ns}}      & $\tau(\mathcal{D})$                        & Total distillation runtime                    & \qty{101}{\mu s}    \\ \midrule
$t$                   & Total algorithm runtime                       & \qty{0.772}{\s}      & $q_\text{dis}$                             & Physical qubits for T-state distillation      & \num{414000}        \\
$q_\text{log}$        & Physical qubits for logical qubits            & \num{190608}         & $q_\text{total}$                           & Total number of physical qubits               & \num{604608}    \\ \bottomrule   
\end{tabular}
\end{table*}

The next step is to find the number of distillation factories $F$ capable of producing the required $M$ T-states during the runtime of the algorithm. First, we use \cref{eq:tstateerror} to calculate the required logical T-state error rate $P_T(\mathcal{D}) = 1.90\cdot 10^{-9}$.
\begin{align}
    M\cdot P_T(\mathcal{D}) = \epsilon_\text{dis} \leq \frac{\epsilon}{3} \label{eq:tstateerror}
\end{align}

The logical output error rate of the distillation units is calculated as in \cref{tab:distillationunits}, using the input logical T-gate error $P_T$ and logical Clifford error $P$. The required output error rate of $1.90\cdot 10^{-9}$ cannot be achieved in one round of distillation with an input T-gate error rate of $0.001$, which means that multiple rounds of distillation will be needed. 

We first consider the last round of distillation. To use the formula for logical output error rate from \cref{tab:distillationunits}:  $P_T(\mathcal{D}) = 35P_T^3+7.1P$, we initially assume that the input T-gate error ($P_T$) does not contribute to the output error rate of the T-factory $P_T(\mathcal{D}_\mathrm{last})$, so that we can calculate the maximum input logical Clifford error rate $P$. We find $P\leq 1.90\cdot 10^{-9} / 7.1 = 2.68 \cdot 10^{-10}$. We use this in \cref{eq:E2} to find a code distance of $\mathcal{D}_\mathrm{last}=17$ for this distillation round, which we use with \cref{eq:Pfromd} to calculate the actual (input) Clifford error rate $P(17) = 3 \cdot 10^{-11}$.
\begin{align}
    P(d) &= a\left(\frac{p}{p^*}\right)^{\frac{d+1}{2}} \label{eq:Pfromd}
\end{align}

 We can then use this to find the maximum allowed input T-state error rate for this round of distillation, using: $35P_T^3 \leq P_T(\mathcal{D}_\mathrm{last}) - 7.1P \rightarrow P_T \leq 3.64\cdot 10^{-4}$. This is the maximum allowed output error for the first distillation round. Many rounds of distillation can be required to reach a certain T-gate error rate, but for this example, we only need two.

\begin{table*}[htb!]
\caption{Characteristics of the distillation units used in the resource estimator. Table adapted from \cite{art:resourceestpaper}, with calculation for time changed to $11\tau(d)$ for consistency with resource estimator output. Characteristics are given for physical and logical space efficient and Reed-Muller (RM) preparation 15-to-1 distillation units.}\label{tab:distillationunits}
\centering
\begin{tabular}{cclll}
\toprule
distillation unit                                                     & \begin{tabular}[c]{@{}l@{}}acceptance probability\end{tabular} & \# qubits & runtime         & \begin{tabular}[c]{@{}l@{}}output error rate\end{tabular} \\ \midrule
\begin{tabular}[c]{@{}c@{}}15-to-1 space eff.\\ physical\end{tabular} & $1-15p_T-356p$                                                    & $12$        & $46t_\text{meas}$ & $p_T(\mathcal{D}) = 35p_T^3+7.1p$                                               \\
\begin{tabular}[c]{@{}c@{}}15-to-1 space eff.\\ logical\end{tabular}  & $1-15P_T-356P$                                                    & $20n(d)$    & $13\tau(d)$  & $P_T(\mathcal{D}) = 35P_T^3+7.1P$                                               \\
\begin{tabular}[c]{@{}c@{}}15-to-1 RM prep.\\ physical\end{tabular}   & $1-15p_T-356p$                                                    & $31$        & $23t_\text{meas}$ & $p_T(\mathcal{D}) = 35p_T^3+7.1p$                                               \\
\begin{tabular}[c]{@{}c@{}}15-to-1 RM prep.\\ logical\end{tabular}    & $1-15P_T-356P$                                                    & $31n(d)$    & \boldmath$11\tau(d)$  & $P_T(\mathcal{D}) = 35P_T^3+7.1P$   \\ \bottomrule
\end{tabular}
\end{table*}
For the first round of distillation, we have an input (physical) T-gate error rate $p_T = 0.001$, and an output (logical) error rate $P_T(\mathcal{D}_1) \leq 3.64\cdot 10^{-4}$. From this, we can calculate the required logical Clifford error rate $P \leq P_T(\mathcal{D}_1)/7.1  = 5.13\cdot 10^{-5}$ and the corresponding code distance $\mathcal{D}_1=5$ as before, using the output error rate in \cref{tab:distillationunits}. The acceptance probability for this type of T-factory is $1-15P_T - 356 P = 0.966$. To get an output of 15 T-states with $99.9\%$ probability, we require 18 distillation units per factory for the first round.  

We can calculate the time required by both rounds of distillation using \cref{tab:distillationunits}. For the first round, we use "space efficient" logical distillation units, and for the second round we use``RM preparation" logical distillation units. This makes the total runtime for both rounds $\tau(\mathcal{D}) = \qty{101}{\mu s}$. Each factory produces $M(\mathcal{D}) = 1$ T-state. 

\begin{align}
F &= \left\lceil\frac{M\cdot \tau(\mathcal{D})}{M(\mathcal{D})\cdot t}\right\rceil \label{eq:E5}
\end{align}
Using \cref{eq:E5}, this gives us $F=23$ T-state distillation factories. The first distillation round uses $18 \cdot 20n(\mathcal{D}_1 = 5) = \num{18000}$ qubits, while the second round of distillation uses $1 \cdot 31 n(\mathcal{D}_\text{last} = 17) = \num{17918}$ qubits. Each factory thus requires $n(\mathcal{D}) = \max(\num{18000}, \num{17918}) = \num{18000}$ qubits%
, and in total $q_\text{dis} = F \cdot n(\mathcal{D}) = \num{414000}$ physical qubits are required for the 23 T-factories.

The total number of physical qubits is thus $q_\text{total} = q_\text{log} + q_\text{dis} = \num{190608} +\num{414000} = \num{604608}$ qubits.

\section{Automated tools for resource estimation} 
\begin{figure}[htb]
    \centering
    \includegraphics[width=0.4\linewidth]{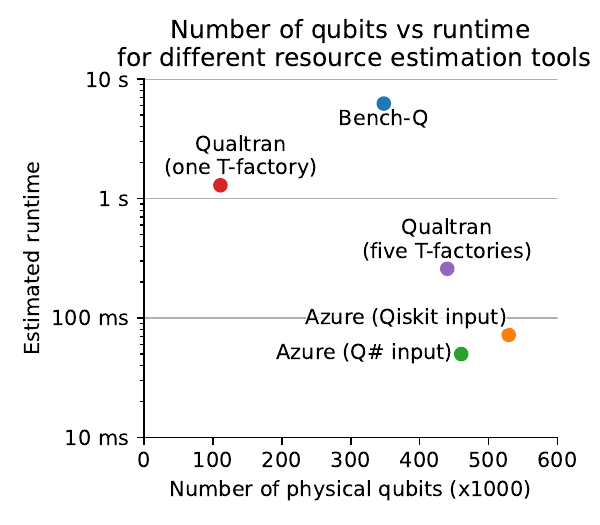}
    \caption{Estimated total number of physical qubits vs computational runtime for one day of operations of the industrial shift scheduling problem defined in \cref{QISS_section}, as estimated by Bench-Q, Qualtran and the AQRE. In all three cases, the built-in models of superconducting qubits were used, with an error rate of $0.001$ and an error budget of $0.001$.}
    \label{fig:different_resource_estimators}
\end{figure}

\label[appendix]{app:automated_res_tools}
We compare three different automated tools to estimate the resources required for a single day of operations in our industrial shift scheduling problem: Qualtran~\cite{misc:qualtran}, Bench-Q~\cite{misc:benchq} and the Azure Quantum Resource Estimator (AQRE,~\cite{art:resourceestpaper}). \cref{fig:different_resource_estimators} shows the number of physical qubits and the total runtime estimated by the three tools. We analyze each case in more detail below. 

\subsection{Qualtran}
Qualtran is a project developed by Google, consisting of a library with quantum abstractions and a rudimentary resource estimator~\cite{misc:qualtran}. It includes qubit, T-state factory and rotation cost models from~\cite{art:resourceestpaper} and~\cite{art:GidneyFowler}. The logical resources of our circuit were counted using the Bloq abstraction, generated from the same QASM code as used for the AQRE and Bench-Q estimates. The resulting counts were used as the input for Qualtran, using the GidneyFowler models~\cite{art:GidneyFowler}, with the \textit{SevenDigitsOfPrecisionConstantCost} rotation cost model and the \textit{simplifiedSurfaceCode} QEC scheme. This gives the following estimation result: \num{111000} physical qubits, and a total runtime of $1.3$ seconds.

\subsection{Bench-Q}
Bench-Q is a resource estimation tool developed by Zapata AI as part of the DARPA Quantum Benchmarking program~\cite{art:zapatabenchq}.
It includes built-in models of trapped-ion and superconducting qubits and supports QASM input through the use of Qiskit. For the results in \cref{fig:different_resource_estimators}, we used a Qiskit implementation of the industrial shift scheduling algorithm described in \cref{QISS_section}, with the "BASIC\_SC\_ARCHITECTURE\_MODEL" and an error budget of $0.001$. Bench-Q outputs an estimate of \num{348000} physical qubits, and a total runtime of $6.2$ seconds.

\subsection{Azure Quantum Resource Estimator}
The AQRE is a cloud-based framework developed by Microsoft that abstracts the layers of a quantum computer stack. It allows the user to customize physical characteristics such as noise parameters, gate and measurement times, the error budget, and the quantum error correction code. The input quantum circuit can be provided in either Q\# or Qiskit~\cite{misc:resestdocumentation}. The AQRE has been used for quantum resource estimation for various applications in~\cite{art:resourceestpaper, art:usingazureresest, art:hansen2023resource}.

The influence of gate decompositions and resource counting can clearly be seen in \cref{fig:different_resource_estimators} in the difference between the estimates obtained using Q\# and Qiskit. 
For a single day of industrial shift scheduling, when using the AQRE with Qiskit as the input language we obtain an estimate of \num{529000} physical qubits, and a runtime of \qty{72}{\ms}. On the other hand, when using Q\#, the AQRE estimates \num{460000} physical qubits, and a runtime of \qty{50}{\ms}. Using Q\# produces more optimal counts for the number of logical resources, because the AQRE uses (abstractions of) optimization techniques through which some operations are implemented with additional measurements to reduce the total required number of T-states when compiling the Q\# circuit.  With fewer T-states, fewer factories and logical cycles are required, leading to fewer physical qubits and a shorter estimated runtime. The AQRE does not use the same optimization techniques on the circuit when it is input using Qiskit, because (integration with) this type of T-state optimization is not currently available. We expect that Qiskit can be used to produce similar results to the Q\# input by tailoring the circuit optimization and gate decompositions to this specific use case. 

In the main text, we use the AQRE with Q\# input for our resource estimations.

\section{Resource estimates  with the scenarios from Beverland et al.}\label[appendix]{app:predefinedscens}

Beverland et al. (2022) use six scenarios in their resource estimation: two based on superconducting qubits, two on Majorana qubits, and two based on trapped-ion qubits~\cite{art:resourceestpaper}. These scenarios are included as predefined scenarios in the AQRE. We will use these scenarios to estimate the quantum resources for our use case. 

\begin{figure}[!hbt]
    \centering
    \begin{subfigure}[t]{0.4\linewidth}
    \centering 
    \includegraphics[width=\linewidth]{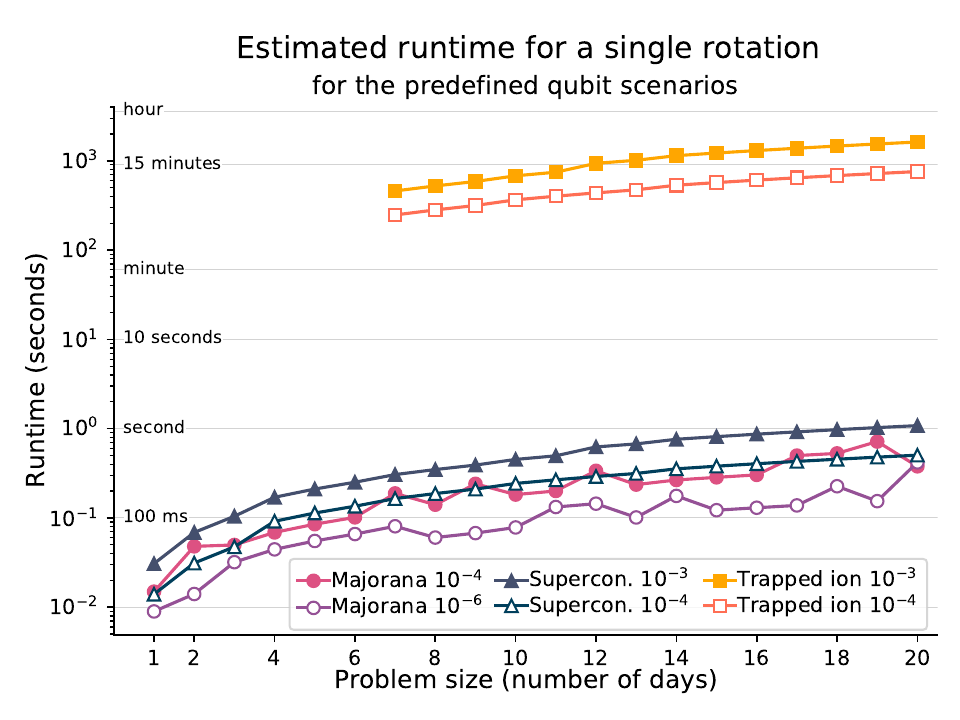}
    \caption{Estimated runtime for a single Grover rotation.} \label{fig:results_predefined_runtime}
    \end{subfigure}
    \hfill
    \begin{subfigure}[t]{0.4\linewidth}
    \centering  
    \includegraphics[width=\linewidth]{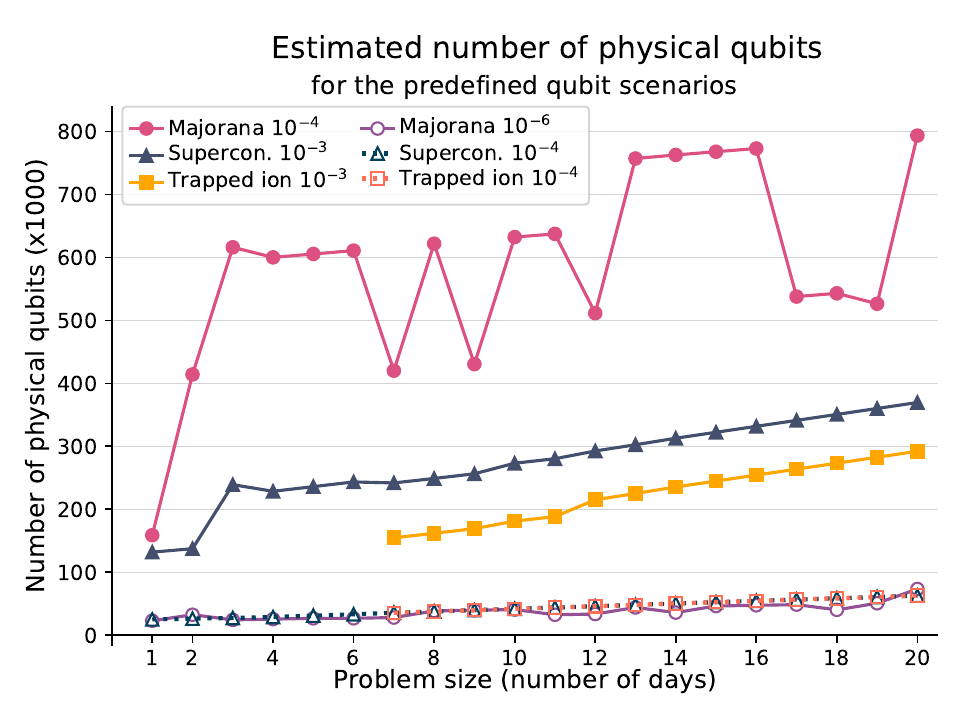}
    \caption{Estimated number of physical qubits required for a single Grover rotation.}\label{fig:results_predefined_qubits}
    \end{subfigure}
    \caption{Resource estimator results for the different scenarios from~\cite{art:resourceestpaper}, for a single Grover rotation with an error budget of 0.25. The runs for trapped-ion qubits start at a problem size of 7 days because of resource estimator limitations. The exact same number of qubits are required for the algorithm for trapped-ion or superconducting qubits with error rates of $10^{4}$.} \label{fig:results_predefined}
\end{figure}

The parameters are summarized in \cref{tab:params_predefined}, and the results are plotted in \cref{fig:results_predefined}. The scenarios based on trapped-ion qubits start from a problem size of 7 days. The computations for smaller problem sizes did not give results because of limitations of the resource estimator when constructing T-factories.

From the six scenarios, using the Majorana qubits with an error rate of $10^{-4}$ requires the highest number of physical qubits. This can be seen clearly in \cref{fig:results_predefined_qubits}. More physical superconducting qubits are required than trapped-ion qubits when the error rate for both is $10^{-3}$, but when the error rate is $10^{-4}$, the exact same number of physical qubits are required for these two technologies. A similar number of physical qubits is required when using Majorana qubits with error rates of $10^{-6}$ or superconducting and trapped-ion qubits with error rates of $10^{-4}$.

Consistent with the results in~\cite{art:resourceestpaper}, trapped-ion qubits are the slowest technology by several orders of magnitude. The runtime required by the Majorana qubits is approximately the same as for the superconducting qubits. This is at odds with the results in~\cite{art:resourceestpaper}, where Majorana qubits are consistently faster than other technologies. This difference could be caused by differences in problem size, error budget or characteristics of the circuit, such as the number of T-gates compared to the circuit length. 

These results validate the use of superconducting qubits for our resource estimations. There are no current Majorana-based quantum systems that we can use to check the validity of the parameter values for the `Majorana' scenarios, 
and the runtime of the trapped-ion qubits means it is unlikely that speedup will be achieved for this application. Superconducting qubits perform consistently well for this application, there are current systems with more than a thousand superconducting qubits~\cite{art:naturecondor,art:atom1180} and roadmaps are available for superconducting systems up to a million qubits.

\begin{table}[!htb]
\caption{Parameters for the resource estimator six qubit scenarios from~\cite{art:resourceestpaper}, of which results are plotted in \cref{fig:results_predefined}.} \label{tab:params_predefined}
\centering
\begin{tabular}{@{}lcccccc@{}}
\toprule
Name                                & \begin{tabular}[c]{@{}c@{}}Majorana\\ $10^{-4}$\end{tabular}       & \begin{tabular}[c]{@{}c@{}}Majorana\\ $10^{-6}$\end{tabular}       & \begin{tabular}[c]{@{}c@{}}Superconducting\\ $10^{-3}$\end{tabular} & \begin{tabular}[c]{@{}l@{}}Superconducting\\ $10^{-4}$\end{tabular} & \begin{tabular}[c]{@{}c@{}}Trapped ion\\ $10^{\shortminus3}$\end{tabular} & \begin{tabular}[c]{@{}c@{}}Trapped ion\\ $10^{\shortminus4}$\end{tabular} \\ \midrule
1Q meas.\ error rate & $10^{-4}$    & $10^{-6}$    & $10^{-3}$   & $10^{-4}$   & $10^{-3}$   & $10^{-4}$   \\
1Q gate error rate   & -       & -       & $10^{-3}$   & $10^{-4}$   & $10^{-3}$   & $10^{-4}$   \\
2Q gate error rate*   & $10^{-4}$   & $10^{-6}$   & $10^{-3}$   & $10^{-4}$   & $10^{-3}$   & $10^{-4}$   \\
T-gate error rate    & $10^{-2}$    & $10^{-2}$    & $10^{-3}$   & $10^{-4}$   & $10^{-6}$   & $10^{-6}$   \\ \midrule
1Q meas.\ time       & \qty{100}{\ns}  & \qty{100}{\ns}  & \qty{100}{\ns} & \qty{100}{\ns} & \qty{100}{\mu s} & \qty{100}{\mu s} \\
1Q gate time         & -       & -       & \qty{50}{\ns}  & \qty{50}{\ns}  & \qty{100}{\mu s} & \qty{100}{\mu s} \\
2Q gate time*         & \qty{100}{\ns} & \qty{100}{\ns} & \qty{50}{\ns}  & \qty{50}{\ns}  & \qty{100}{\mu s} & \qty{100}{\mu s} \\
T-gate time          & \qty{100}{\ns}  & \qty{100}{\ns}  & \qty{50}{\ns}  & \qty{50}{\ns}  & \qty{100}{\mu s} & \qty{100}{\mu s} \\ 
\bottomrule
\multicolumn{7}{l}{* Two-qubit joint measurements for the Majorana qubits.}
\end{tabular} 
\end{table}

\section{Choice of characteristics of the high-fidelity qubit scenarios} \label[appendix]{app:futurequbitdetail}

In this appendix, we provide more detail about our choice of characteristics for our high-fidelity qubit scenarios, which are found in \cref{tab:paramse3_e9}. The estimation results are plotted in \cref{fig:results_e3_e9}.

\begin{table}[!htb] 
\caption{Parameters for the resource estimator for high-fidelity superconducting qubits plotted in \cref{fig:results_e3_e9,fig:physical_qs_v_runtime12}. }\label{tab:paramse3_e9}
\centering
\setlength{\tabcolsep}{6pt}
\begin{tabular}{lcccrrc}
\toprule
Name                & $10^{-3}$   & $10^{-4}$   & $10^{-6}$   & $10^{-8}$   & $10^{-9}$ & perfect\\ \midrule
1Q meas.\ error rate & $10^{-3}$  & $10^{-4}$  & $10^{-6}$  & $10^{-8}$  & $10^{-9}$ & $0$\\
1Q gate error rate  & $10^{-3}$  & $10^{-4}$  & $10^{-6}$  & $10^{-8}$  & $10^{-9}$ & $0$\\
2Q gate error rate  & $10^{-3}$  & $10^{-4}$  & $10^{-6}$  & $10^{-8}$ & $10^{-9}$ & $0$\\ 
T-gate error rate   & $10^{-3}$  & $10^{-4}$  & $10^{-6}$  &\textbf{$\mathbf{5\times 10^{-8}}$} &\textbf{$\mathbf{5\times 10^{-8}}$} & $0$\\
Idle error rate     & $10^{-3}$  & $10^{-4}$  & $10^{-6}$  & $10^{-8}$  &$10^{-9}$ & $0$\\ \midrule
\begin{tabular}{@{}l@{}}
1Q meas.\ time \\ 1Q gate time \\ 2Q gate time \\ T-gate time \end{tabular}      & \begin{tabular}{r} \qty{100}{\ns} \\ \qty{50}{\ns}  \\ \qty{50}{\ns}  \\ \qty{50}{\ns} \end{tabular} & \begin{tabular}{r} \qty{100}{\ns} \\ \qty{50}{\ns}  \\ \qty{50}{\ns}  \\ \qty{50}{\ns} \end{tabular} & \begin{tabular}{r} \qty{10}{\ns} \\ \qty{5}{\ns}  \\ \qty{5}{\ns}  \\ \qty{5}{\ns} \end{tabular}  & \begin{tabular}{r} \qty{10}{\ns} \\ \qty{5}{\ns}  \\ \qty{5}{\ns}  \\ \qty{5}{\ns} \end{tabular}  & \begin{tabular}{@{}r@{}} \qty{1}{\ns} \\ \qty{0.5}{\ns}  \\ \qty{0.5}{\ns}  \\ \qty{0.5}{\ns} \end{tabular}   & \begin{tabular}{@{}c@{}} \qty{0.2}{\ns} \\ --  \\ -- \\ -- \end{tabular} \\ \bottomrule
\end{tabular}
\end{table}

\cref{tab:paramse3_e9} shows the error rates for the qubits used in the estimates plotted in \cref{fig:physical_qs_v_runtime12}. The gate times are scaled according to the measurement time in the plot: in the first set, the measurement operation takes \qty{1}{\ns} and gate operations take \qty{0.5}{\ns}, in the second set, measurement takes \qty{10}{\ns} and gates take \qty{5}{\ns}. In the third set, measurement takes \qty{100}{\ns} and gate operations take \qty{50}{\ns}.

For the scenarios for superconducting qubits that are plotted in \cref{fig:results_e3_e9}, we use different measurement times for different error rates. The superconducting qubits used in previous resource estimations have flat error rates of $10^{-3}$ and $10^{-4}$~\cite{art:resourceestpaper} and gate and measurement times of \qty{50}{\ns} and \qty{100}{\ns}, respectively. Based on this we set the gate/measurement time for the scenarios with error rates of $10^{-6}$ and $10^{-8}$ to $5$/\qty{10}{\ns} and for the scenario with error rate of $10^{-9}$ the gate/measurement time will be $0.5$/\qty{1}{\ns}. 

The AQRE can only be used for qubits with error rates bigger than zero. To get estimates for perfect qubits, we used the estimated logical qubits and the logical depth as output by the AQRE. To get the runtime, the logical depth was multiplied by a value of \qty{0.2}{\ns}. This represents a quantum computer with a clock frequency of \qty{5}{\GHz}, that does not use any error correcting methods but does take some mapping and gate decompositions into account. 

When the gate error rate is below a certain limit compared to the required logical T-gate error rate, the AQRE cannot calculate the number and type of T-factories. 
To circumvent this, we set the minimum T-gate error to $5\cdot 10^{-8}$. Even with this, no estimates could be made for problem sizes below 4 days for the scenarios with error rates of $10^{-8}$ and $10^{-9}$, and below 2 days for the scenario with error rate of $10^{-6}$. These are shown as missing data points in \cref{fig:results_e3_e9}.

For the brute-force search, we set the time required for a checking a single guess at \qty{1}{\ns}. This is much shorter than it takes for an actual guess for the best solution, which includes calculations and requirement checks. The value of \qty{1}{\ns} was chosen to represent a parallel implementation of the search, where a new guess is completed (on average) every nanosecond. How long an actual brute-force search takes depends on many factors, including quality of implementation and resources available for parallel execution. The runtime plotted here serves as an indication for the exponentially increasing classical execution time. 

When decreasing gate error rates to $10^{-8}$ or $10^{-9}$, we reach the limits of what the AQRE can be used for. At these error rates, the AQRE cannot calculate T-factories. Decreasing the error rate for T-gates further is not possible, and the estimation for the number of physical qubits required is most likely fully determined by the T-gate error and factors that we have no influence over. It is likely that a real quantum computer with such low qubit error rates requires less qubits than are estimated here and has other limiting factors that we cannot predict from current technologies.

\section{Error budgets} \label[appendix]{app:errorbudget}
We use the error budget to scale our estimates for running QISS until a solution is found, which requires ${\sim}\sqrt{N}$ applications of the Grover rotation operator. We will first discuss the influence of the error budget, and then explain our technique for obtaining our estimates for $\sqrt{N}$ Grover rotations.

\subsection{Division of the error budget} \label[appendix]{appsec:errorbudgetdiv}

The error budget for logical qubits is the fraction of runs that are allowed to fail from an error in one the of logical qubits. A lower logical qubit error budget lowers the maximum allowed error rate for each logical qubit, which increases the required code distance and therefore the number of physical qubits.

A lower error budget for rotation gate synthesis means an increased number of T-states required for the decomposition of each rotation gate~\cite{art:kitaev}, and thereby the total number of T-states required. This tends to increase the number of T-state factories and therefore the total number of physical qubits and/or the runtime of the algorithm.

A lower error budget for T-state distillation means a lower maximum error for the distillation of each T-state. This influences the choice of the type of T-factory, and can mean an increase in code distance per round of distillation, an increase in the number of distillation rounds, or both. 

By dividing our error budget equally, we do not take into account the runs which fail due to more than one error. Our effective error budget is thus slightly lower, at $(1-(1-0.25/3)^3)\approx 23\%$. This equal division might not be the most optimal division of the error budget, since all three parts are unlikely to contribute equally to the final estimated resources. The optimal division will depend on all factors of the resource estimation, and is outside of the scope of this paper. 

\subsection{Extrapolating from a single iteration to \texorpdfstring{$\sqrt{N}$}{sqrt(N)} iterations} \label[appendix]{app:scalingtosqrtN}

For full runs of Grover Adaptive Search (GAS) we need to make estimates for the resources required to run $\sqrt{N}$ iterations of Grover's search. 
Due to limitations in the circuit size that we were able to submit to the AQRE, we use the following method to obtain the resources required for all $\sqrt{N}$ rotations of Grover's search.

For each loop of GAS, we run a circuit with a randomly determined number of operations of the Grover rotation operator. If we want to have the same error budget for each circuit, then for longer circuits the error budget will be lower per logical gate and per logical qubit. This means more logical time steps and more physical qubits are required for these longer circuits. 

The number of applications of the Grover rotation operator per loop of GAS is between zero and a number $m$, which is increased every time the loop does not result in an improved solution. Every time it does find an improved solution, $m$ is reset to $1$~\cite{art:krol2024qiss}. This means that within the total of $\sqrt{N}$ iterations, we are often executing circuits with a low number of iterations, and not as many longer circuits. That the algorithm will land on a single circuit containing $\sqrt{N}$ applications of the  Grover rotation operator will be rare. Doing estimates based on this longest possible circuit will give an upper bound, but in reality most runs will be for much shorter circuits. The lower bound will be given by multiplying estimates for a circuit with a single Grover rotation by the total allowed number of rotations, $\sqrt{N}$. This would be the number of resources required for a run of GAS where only circuits with one Grover rotation are selected, which might be the case if the maximum allowed number of iterations $m$ is reset to $1$ (almost) every loop.

For getting an upper bound for the number of resources required by our algorithm, we need to run resource estimates for circuits with $\sqrt{N}$ applications of the Grover operator. Because of AQRE limits, we can only do that directly for problem sizes up to 7 days ($\sqrt{N} = \num{16384}$)\footnote{Runs for problem sizes bigger than this number gave the following error: "It took longer than the max timeout of 10 minutes to execute one of the steps of resources estimation, so the job was stopped."}. For bigger problem sizes, we instead use an approximation that we now explain. 

The simplest way to make such approximations is to do the resource estimation for a single application of the Grover rotation operator for a given problem size, but reduce the error budget for this single iteration $\epsilon_1$ such that the combined error budget for $n$ iterations $\epsilon_n$ is equal to the error budget of a circuit that contains all $\sqrt{N}$ iterations. This scaling is shown in \cref{eq:e_n}.

\begin{align}
    \epsilon(n) &= n \cdot \epsilon_1 = \epsilon_n \label{eq:e_n}\\
    C(n, \epsilon_n) &\approx n(M_\text{meas} + M_\text{R} + M_\text{T} + 3M_\text{Tof}) + \left\lceil A \log_2\left(\frac{n M_\text{R}}{\epsilon_1 n/3}\right) + B \right\rceil n D_\text{R} = n C(1, \epsilon(n)) \label{eq:C_n} \\
    M(n, \epsilon_n) &\approx \left\lceil A \log_2\left(\frac{n \cdot M_\text{R}}{n \cdot \epsilon_1/3}\right) + B \right\rceil n M_\text{R} + n( 4 M_\text{Tof} + M_\text{T}) = n M(1, \epsilon(n))\label{eq:M_n} \\
    d(n, \epsilon_n) &\approx \left\lceil \frac{2 \log(3\cdot a\cdot Q\cdot n\cdot C(1, \epsilon(n))/(n \cdot \epsilon_1)}{\log(p^*/p)} -1 \right\rceil_{odd} = d(1, \epsilon(n)) \label{eq:d_n}
\end{align}

From this follow \cref{eq:C_n,eq:M_n,eq:d_n}, which are modified from the equations in \cite{art:resourceestpaper} or \cref{app:analyticalexample}.
The code distance for the T-factories is essentially the same as for the logical qubits, using \cref{eq:d_n} with $n\cdot M(1, \epsilon(n))$ instead of $n\cdot C(1, \epsilon(n)$. The rest of the estimation results follow from the results of these equations.

The two main resources we are interested in are runtime and number of qubits. For a set problem size, the estimation for the number of qubits depends only on the number of qubits required per logical qubit, and the number of qubits required for the T-factories. The first is purely dependent on the minimum code distance to achieve the required logical qubit error rate. We assume that for circuits with the same code distance, the same number and type of T-factories are selected. We further assume that the setup at the start and the measurements at the end of the circuit do not contribute much to the total circuit depth or to the error budget, so that the gate counts and depth of a circuit with $n$ iterations are equal to $n$ times a circuit with one application of the Grover rotation operator. And finally, we assume that the probability of errors occurring in an iteration is independent of the other iterations, and that the error budget is divided equally over all iterations.

\begin{figure}[hbt]
    \centering
    \begin{subfigure}[t]{0.4\linewidth}
    \centering 
    \includegraphics[width=\linewidth]{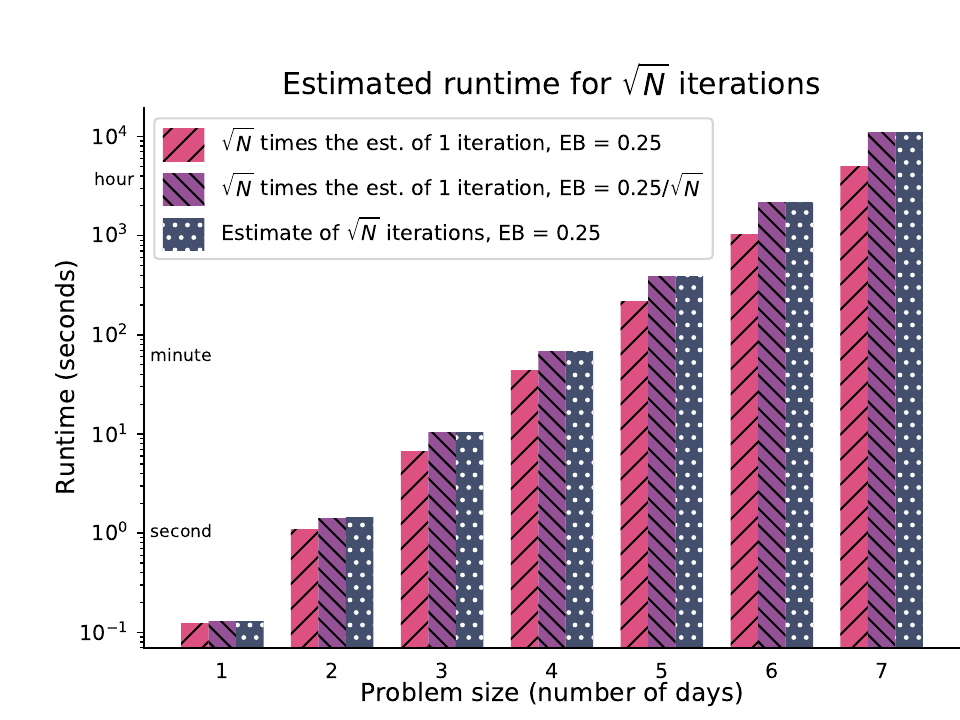}
    \caption{Estimated runtime for $\sqrt{N}$ Grover rotations.}
    \label{fig:estimated_runtime_linearscaling}
    \end{subfigure}
    \hspace{0.05\linewidth}
    \begin{subfigure}[t]{0.4\linewidth}
    \centering  
    \includegraphics[width=\linewidth]{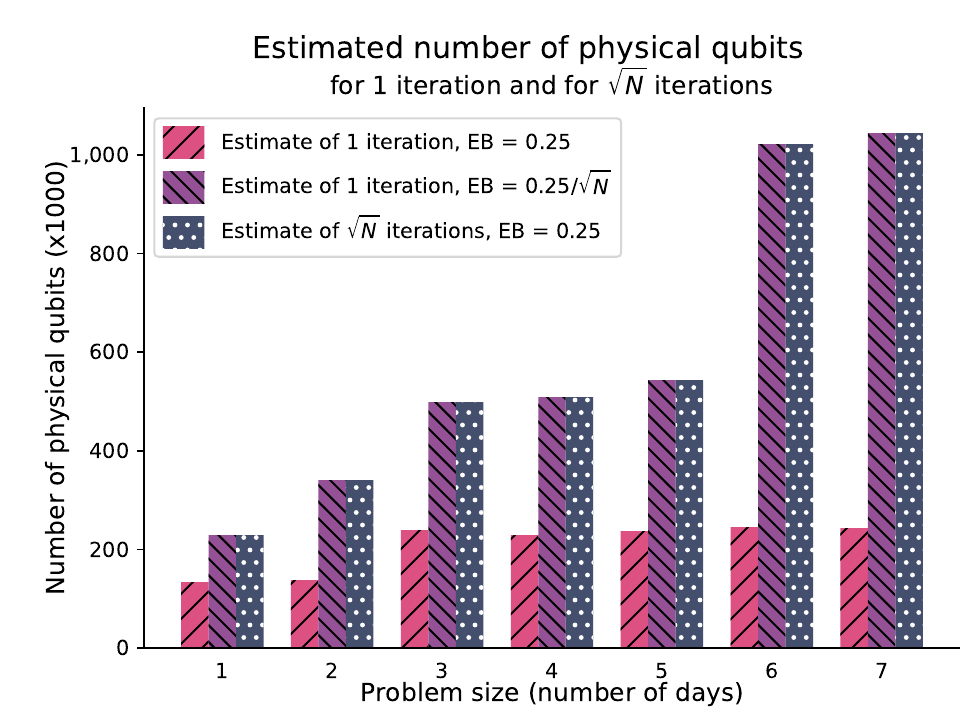}
    \caption{Estimated number of physical qubits for $\sqrt{N}$ Grover rotations.}
    \label{fig:estimated_phys_linearscaling}
    \end{subfigure}
    \caption{Resource estimation results for different methods of error budget (EB) scaling, for problem sizes from one to seven days. The estimates are based on resource estimator results for the superconducting qubits with a flat error rate of $10^{-3}$ (see \cref{tab:paramsupercond}). Estimates were made using the scenario for superconducting qubits with a flat error rate of $10^{-3}$ (see \cref{tab:paramsupercond}), for the QISS algorithm~\cite{art:krol2024qiss} with a single Grover rotation and a constant error budget of $0.25$, for a single Grover rotation and a scaled error budget, calculated as $0.25/\sqrt{N}$, and for the algorithm containing $\sqrt{N}$ Grover rotations with a constant error budget of $0.25$. To get estimates for $\sqrt{N}$ applications of the Grover rotation operator, the runtime results from a single rotation were multiplied by $\sqrt{N}$. Qubit number estimates are not multiplied, the difference between $1$ and $\sqrt{N}$ iterations is caused only by the difference in error budget.} \label{fig:results_linearscaling}
\end{figure}

To check the influence of these assumptions, estimations were made up to a problem size of seven days. The results of these can be found in \cref{fig:results_linearscaling}. As is clear from the plots, the estimations using the scaled error budget correspond closely to the estimates for the circuits with $\sqrt{N}$ iterations. Therefore, this is the method we will be using for all our experiments when the required circuit length exceeds the limits of the AQRE.

\clearpage
\section{Data} 
\begin{table}[!h]
\caption{Data used to generate \cref{fig:quantumroadmap}} \label{tab:quantumroadmap}
\centering
\begin{tabular}{@{}lccccccl@{}}
\toprule
Company        & Modality        & Designation      & Year    & \# of qubits & Released? & Source                                             & Note                                    \\ \midrule
IBM            & Superconducting & Canary           & 2017    & \num{5}            & yes     & \cite{art:IBMnewroadmap}                           &                                         \\
               &                 & Albatross        & 2018    & \num{16}           & yes     & \cite{art:IBMnewroadmap}                           &                                         \\
               &                 & Penguin          & 2019 & \num{20}           & yes     & \cite{art:IBMnewroadmap}                           &                                         \\
               &                 & Falcon           & 2020    & \num{27}           & yes     & \cite{art:IBMnewroadmap}                           &                                         \\
               &                 & Eagle            & 2021 Q4 & \num{127}          & yes     & \cite{art:IBMnewroadmap}                           &                                         \\
               &                 & Osprey           & 2022 Q4 & \num{433}          & yes     & \cite{art:IBMnewroadmap}                           &                                         \\
               &                 & Condor           & 2023 Q4 & \num{1121}         & yes     & \cite{art:naturecondor}                            &                                         \\
               &                 & Kookaburra       & 2025    & \num{4158}         &         & \cite{art:IBMnewroadmap}                            & \# of qubits from IBM's previous roadmap~\cite{art:IBMoldroadmap}                      \\
               &                 & Cockatoo         & 2027    &              &         & \cite{art:IBMnewroadmap}                           & \# of qubits is not specified           \\ 
               &                 & Blue Jay         & 2033    & \num{100000}         &         & \cite{art:IBMnewroadmap}                           & Also listed as \num{2000} (logical) qubits    \\ \midrule
IBM            & Superconducting & Heron            & 2023 Q4 & \num{133}          & yes     & \cite{art:IBMnewroadmap}                           &                                         \\
modular        &                 & Flamingo         & 2024    & \num{156}          &         & \cite{art:IBMnewroadmap}                           &                                         \\
               &                 & Crossbill        & 2024    & \num{408}          &         & \cite{art:IBMoldroadmap}                           &                                         \\
               &                 & Heron            & 2024    & \num{399}          &         & \cite{art:IBMnewroadmap}                           & $133\times 3$                                   \\
               &                 & Flamingo         & 2025    & \num{1092}         &         & \cite{art:IBMnewroadmap}                           & $156\times 7$                                   \\
               &                 & Starling         & 2029    & \num{200}          &         & \cite{art:IBMnewroadmap}                           &                                         \\
               &                 & Blue Jay         & 2033    & \num{100000}         &         & \cite{art:IBMnewroadmap}                           & Also listed as \num{2000} (logical) qubits                                        \\ \midrule
Google         & Superconducting & Foxtail          & 2017    & \num{22}           & yes     & \cite{art:googlesycamore}                          &                                         \\
               &                 & Brislecone       & 2018 Q2 & \num{72}           & yes     & \cite{art:googlesycamore}                          &                                         \\
               &                 & Sycamore         & 2019 Q3 & \num{53}           & yes     & \cite{art:googlesycamore}  & Has 53 ‘effective’ qubits               \\
               &                 & M3               & 2025+    & \num{1000}         &         & \cite{art:googleroadmap}                           &                    \\
               &                 & M4               & -   & \num{10000}        &         & \cite{art:googleroadmap}                           &                  \\
               &                 & M5               & -    & \num{100000}       &        & \cite{art:googleroadmap}                           &                   \\
               &                 & M6               & -    & \num{1000000}      &         & \cite{art:googleroadmap}                           &                   \\ \midrule
Rigetti        & Superconducting & Agave            & 2017 Q2 & \num{8}            & yes     & \cite{art:rigettibuilt}                            &                                         \\
               &                 & Acorn            & 2017 Q4 & \num{19}           & yes     & \cite{art:rigettibuilt}                            &                                         \\
               &                 & Aspen-1          & 2018 Q4 & \num{16}           & yes     & \cite{art:rigettibuilt}                            &                                         \\
               &                 & Aspen-4          & 2019 Q1 & \num{13}           & yes     & \cite{art:rigettibuilt}                            & Higher fidelity than previous iteration \\
               &                 & Aspen-7          & 2019 Q4 & \num{28}           & yes     & \cite{art:rigettibuilt}                            &                                         \\
               &                 & Aspen-9          & 2021 Q1 & \num{32}           & yes     & \cite{art:rigettibuilt}                            &                                         \\
               &                 & Aspen-11         & 2021 Q4 & \num{40}           & yes     & \cite{art:rigettibuilt}                            &                                         \\
               &                 & Aspen-M-1        & 2022 Q1 & \num{80}           & yes     & \cite{art:rigettibuilt}                            &                                         \\
               &                 & Ankaa-2          & 2024 Q1 & \num{84}           & yes     & \cite{art:rigettiannounces}                        &                                         \\
               &                 & Lyra             & 2025    & \num{336}          &         & \cite{art:rigettiroadmap}                          &                                         \\
               &                 & -                & 2026    & \num{1000}         &         & \cite{art:rigettiroadmap}                          &                                         \\
               &                 & -                & 2028    & \num{4000}         &         & \cite{art:rigettiroadmap}                          &                                         \\ \midrule
Oxford Quantum & Superconducting & Lucy             & 2022    & \num{8}            & yes     & \cite{art:oqclucy}                                 &                                         \\
Circuits       &                 & OQC Toshiko      & 2023 Q4 & \num{32}           & yes     & \cite{art:oqctoshiko}                              &                                         \\ \midrule
Intel          & Superconducting & -                & 2017 Q3 & \num{17}           & yes     & \cite{art:intel2017}                               &                                         \\
               &                 & Tangle lake      & 2018 Q1 & \num{49}           & yes     & \cite{art:intel2018}                               &                                         \\
               & Quantum dots    & Tunnel falls     & 2023 Q3 & \num{12}           & yes     & \cite{art:intelquantumdots}                        &                                         \\ \midrule
IonQ           & Trapped ions    & Harmony          & 2019    & \num{11}           & yes     & \cite{art:ionqpast}                                &                                         \\
               &                 & Aria             & 2022    & \num{21}           & yes     & \cite{art:ionqaria}                                &                                         \\
               &                 & Forte            & 2023    & \num{32}           & yes     & \cite{art:ionqpast}                                &                                         \\
               &                 & Forte enterprise & 2024    & \num{35}           &         & \cite{art:ionqannounces}                           &                                         \\
               &                 & Tempo            & 2025    & \num{64}           &         & \cite{art:ionqannounces}                           &                                         \\
               &                 & -                & 2026    & \num{256}          &         & \cite{art:ionqpast}                                &                                         \\
               &                 & -                & 2027    & \num{384}          &         & \cite{art:ionqpast}                                &                                         \\
               &                 & -                & 2028    & \num{1024}         &         & \cite{art:ionqpast}                                &                                         \\ \midrule
Quantinuum     & Trapped ions    & H1               & 2020    & \num{20}           & yes     & \cite{art:quantinuumhardware}                      &                                         \\
               &                 & H2-1               & 2023 Q2 & \num{32}           & yes     & \cite{art:quantinuumhardware}                      &                                         \\ 
               &                 & H2-2             & 2024 Q2 & \num{56}           & yes     & \cite{art:quantinuumH22}                           & \\\midrule
Xanadu         & Photonic        & X2               & 2018    & \num{2}            & yes     & \cite{art:xanadupast}                              &                                         \\
               &                 & X4               & 2019    & \num{4}            & yes     & \cite{art:xanadupast}                              &                                         \\
               &                 & X8               & 2020 Q4 & \num{8}            & yes     & \cite{art:xanadupast}                              &                                         \\
               &                 & X12              & 2020 Q4 & \num{12}           & yes     & \cite{art:xanadux12}                               &                                         \\
               &                 & Borealis         & 2022 Q2 & \num{216}          & yes     & \cite{art:xanaduborealis}                          &                                         \\ \midrule
Atom computing & Neutral atoms   & Phoenix          & 2021 Q3 & \num{100}          & yes     & \cite{art:atomphoenix}                             &                                         \\
               &                 & -                & 2024    & \num{1180}         &         & \cite{art:atom1180}                                &                                         \\ \midrule
QuEra          & Neutral atoms   & Aquila           & 2023 Q3 & \num{256}          & yes     & \cite{art:queraaquila}                             &                                         \\
               &                 & -                & 2024    & '\textgreater \num{256}'         &         & \cite{art:queraroadmap}                            & Not plotted                             \\
               &                 & -                & 2025    & '\textgreater \num{3000}'        &         & \cite{art:queraroadmap}                            & Plotted as \num{3000} qubits                  \\
               &                 & -                & 2026    & '\textgreater \num{10000}'       &         & \cite{art:queraroadmap}                            & Plotted as \num{10000} qubits                 \\ \bottomrule
\end{tabular}
\end{table}

\printbibliography
\end{refsection}
\end{document}